\documentclass[12pt]{article}

\usepackage{fancyhdr}
\usepackage{epsfig}
\usepackage{cite}
\usepackage{graphicx}
\usepackage[centertags]{amsmath}
\usepackage{theorem}

\usepackage{youngtab}
\usepackage{graphicx}
\usepackage{amssymb}
\usepackage{latexsym}
\usepackage{epic}
\usepackage{pstricks}
\usepackage{color}
\usepackage{mathrsfs}
\usepackage{latexsym}
\usepackage{cancel}
\usepackage{epic}
\usepackage{braket}
\usepackage{slashed}
\newcommand{\be}{\begin{equation}}
\newcommand{\ee}{\end{equation}}
\newcommand{\bea}{\begin{eqnarray}}

\newcommand{\eea}{\end{eqnarray}}

\numberwithin{equation}{section}
\def\hybrid{\topmargin 22pt    \oddsidemargin 0pt 
      \headheight 0pt \headsep 0pt
      \textwidth 6.5in        
       \textheight 9in         
      \marginparwidth .875in
      \parskip 5pt plus 1pt   \jot = 1.5ex}

\hybrid

\usepackage{caption} 
\newcommand{\refe}[1]{Eqn.~(\ref{#1})}

\renewcommand{\thefootnote}{\fnsymbol{footnote}}

\begin{document}
\center{
\begin{flushright} 
DESY 12-171\\
10.1007/JHEP02(2013)132
\end{flushright}
\vspace{1cm}
\begin{center}
{\Large\textbf{Holography for General Gauge Mediation}}
\vspace{1cm}

\textbf{Moritz McGarrie$^{1,}$\footnote[2]{\texttt{moritz.mcgarrie@desy.de} } 
}\\
\end{center}

{\it{ ${}^1$ 
Deutsches Elektronen-Synchrotron,\\ DESY, Notkestrasse 85, 22607 Hamburg, Germany
}\\
\vspace{0.5cm}

}

\setcounter{footnote}{0}
\renewcommand{\thefootnote}{\arabic{footnote}}

\abstract{We construct a holographic model of spontaneous supersymmetry breaking in analogy to AdS/QCD models.  Integrating out the bulk theory one obtains an entirely four dimensional effective action that encodes spontaneous supersymmetry breaking effects, coupled to external sources.  Using only this four dimensional action it is possible to compute soft masses, scattering cross sections and determine the form factors of vector mesons. This construction lends itself to a more natural comparison with operator product expansions.  }

\section{Introduction}
The  AdS/CFT correspondence   \cite{Maldacena:1997re,Witten:1998qj,Gubser:1998bc,Aharony:1999ti}    has enabled the development of new insights into the problem of strongly coupled gauge theories.   In particular the AdS/QCD programme  \cite{Son:2003et,DaRold:2005ju,Erlich:2005qh,Karch:2006pv}  has developed ``bottom-up'' models that apply the ideas and perspectives of  gauge/gravity duality to QCD.
  Inspired by these we wish to explore AdS/\cancel{SUSY} models as initiated in \cite{Goldberger:2002pc,Chacko:2003tf,Nomura:2004zs,Benini:2009ff,McGuirk:2009am,Abel:2010uw,Abel:2010vb,McGarrie:2010yk,McGarrie:2011av,Fischler:2011xd,McGarrie:2012ks}.   Similar to AdS/QCD, the idea is to construct a setup which attempts to describe a large $N_c$ SQCD model with a spontaneous supersymmetry breaking sector.   In particular it is hoped that the breaking sector will be somewhat similar to ISS constructions \cite{Intriligator:2006dd} in that supersymmetry is broken dynamically and is metastable.  The notation of general gauge mediation (GGM) \cite{Meade:2008wd} is used as a method to encode the supersymmetry breaking effects which are then mediated to the supersymmetric standard model which exists entirely outside the AdS system (on the UV brane). 

The main objectives in using this approach, versus the five dimensional picture, are:
\begin{enumerate}
\item  Using the holographic basis and operator field correspondence one arrives at an entirely four dimensional action that encodes the supersymmetry breaking and the approximate CFT effects without reference to the fifth dimension.  These effects appear as correlators, with dressed vertices, coupled to external sources. This is achieved in  \refe{effectiveaction3}.  We hope that in doing this one can bring warped models of supersymmetry breaking  \cite{Chacko:2003tf,Nomura:2004zs,Benini:2009ff,McGuirk:2009am,Abel:2010uw,Abel:2010vb,McGarrie:2010yk,McGarrie:2011av,McGarrie:2012ks,Bouchart:2011va,Yamada:2011dh,Okada:2011ed} closer to the original GGM proposal   \cite{Meade:2008wd}.
\item  As this notation is more standard within the  AdS/QCD framework, one can make a closer connection to scattering cross sections  of visible $\rightarrow$ hidden processes \cite{Fortin:2011ad}: form factors and operator product expansions arise quite naturally in this basis.  This may be a useful tool to learn more about relating the strongly coupled supersymmetry breaking sector to its UV complete description.  A main result in this regard is \refe{scattering1}.  Some scaling features in common with quark-hadron duality are observed, as a result of the infinite tower of vector mesons between visible and hidden sector states.
\end{enumerate}

It is generally believed that if one builds a consistent supergravity theory on an asymptotically AdS space then it will act as an effective description corresponding to some approximately conformal gauge theory defined on the boundary of that space. In particular one no longer requires an exact duality as in the AdS/CFT correspondence, and although quantifying the discrepency may be difficult, at least in principle one can argue that for momenta $k^2<\Lambda^2$, relative to the effective cutoff,  the two theories should match.  In particular there is an S matrix (due to the UV and IR boundaries that regulate the space) and so particle spectra on the AdS side should correspond to composite states of the approximate CFT.    This  correspondence is used to build a model that encodes an O'Raifeartaigh type messenger model located on the IR brane and which encodes vector meson resonances in the bulk associated with a weakly gauged flavour symmetry of the approximate CFT.

The model is as follows: we assume a large $N_c$ approximately conformal SQCD sector that breaks supersymmetry and admits a supergravity dual  \cite{ArkaniHamed:2000ds,Rattazzi:2000hs,Gherghetta:2010cj,Gherghetta:2006ha}.  Working in the limit that gravity corrections can be ignored and simply supplies the background metric, it is assumed that the approximate CFT has a weakly gauged $SU(N_f)$ flavour symmetry.  On the AdS side this corresponds to a gauge theory in the bulk with dynamical sources.  This $SU(N_f)$ will be associated with the standard model gauge groups.   The standard model gauge groups correspond to $\mathcal{N}=1$ super Yang Mills on the boundary and supply the UV boundary sources  in a gauge vector multiplet $V^0$.  These couple to CFT operators which are dual to a 5d bulk gauge theory with $\mathcal{N}=1$ supersymmetry, regulated by two branes at $z=L_0$ and $z=L_1$ in the $z$ direction.

In a five dimensional theory there are two possible bases: the Kaluza Klein  (KK) basis and the holographic basis.  The poles in the holographic and KK basis necessarily match.    In the KK basis the vector superfield $V_{bulk}$ has a KK expansion such that $A^{\mu}(x,z)=\sum_n A^{\mu}_n(x) f_n(z)$ , with $z$ the fifth directon and in the holographic case one takes $A^{\mu}(q,z)=A^{\mu}_0(q)V(q,z)$ which now takes on a holographic basis, $x$ and $q$ both being four dimensional.  A comparison of these may be found in \cite{Davoudiasl:2009cd,Ponton:2012bi}.  As discussed in \cite{Falkowski:2008yr} the holographic basis is more useful when exploring soft wall models or models where rather than a discrete basis, a continuum basis with mass gap is more appropriate, as may arise from approximately conformal theories.  These types of extensions motivate studying these models in this holographic basis.  Essentially one should use the holographic and Kaluza-Klein basis concurrently, for instance the KK expansion is convenient to understand couplings between different resonances and is most relevant for collider phenomenology as physical mass eigenstates are produced in collisions.  There can be a procedural difference however, in the holographic picture one often wishes to integrate out the bulk dynamics and arrive at an effective Lagrangian on the four dimensional boundary.  Then one uses the boundary action to generate purely four dimensional diagrams.  This is the approach taken here.

This model is based on a slice of $AdS_5$, which is particularly convenient as many things can be solved exactly at leading order. It is found that the conditions of supersymmetry of the Lagrangian will fix the boundary actions and completely determine all the field profiles of the bulk gauge theory.  This indicates that to implement spontaneous supersymmetry breaking one may introduce new degrees of freedom either in the bulk or on the IR brane. 

Before proceeding we wish to summarise a few of the key new results of this paper: integrating out the bulk action one is able to show a four dimensional supersymmetric action on the UV boundary that arises at order $O(N_c)$. In addition one arrives at a separate \emph{spontaneous} supersymmetry breaking effective action on the UV boundary.   Using this effective action,  subleading diagrams may be constructed.   This same effective action may be used to compute scattering cross sections of visible $\rightarrow$ hidden messengers which arise at $O(N^0_c)$.  In addition form factors for  cross sections are identified as a sum of monopole contributions of an infinite tower of vector mesons \emph{directly} analogous to that of AdS/QCD models.  This is related to the non-normalisable mode with Neumann boundary conditions extending from the UV to IR brane.  

In addition to the new results a few issues are clarified.  In particular the relationship between holographic gauge mediation and warped gauge mediation models is explained. It is shown how computations of the beta function fit into this framework and useful ways in which this approach may be extended are given.  We also comment on some overall successes of this framework: supersymmetry is broken spontaneously with \emph{both} gaugino and scalar soft masses that arise naturally as well as a controlled logarithmic running of the beta function allowing for a viable phenomenology.  In addition cross sections are calculable and their form factors match results of the AdS/QCD programme.

In this setup there are three sets of current couplings that need to be specified \refe{1}, \refe{2} and \refe{3}.  The first is a  coupling of the 
supersymmetric standard model gauge fields to the SSM currents
\be
\mathcal{L}\supset   \int d^4 x\  g_{SM}\int d^4 \theta  J_{SM} V_{SM}. \label{1}
\ee
It is typical to work at energy scales $E\sim \hat{M}$ where $\hat{M}=a(L_1) M$ (using \refe{aa}) is the typical mass scale of the supersymmetry breaking sector.  The supersymmetric standard model exists outside the AdS system.  Therefore one should should equate the $V_{SM}$ to a set of source fields $V_0$ on the AdS boundary.   Note that UV localised states such as those contained in $J_{SM}$ are not part of the approximate CFT and only couple to the CFT indirectly through the source fields.

Next one defines a coupling between the boundary sources $V_0$ and the bulk fields
\be
\mathcal{L}\supset \int d^4 x \  g_{SM}\int d^4 \vartheta  \mathcal{O}V_{0}.  \label{2}
\ee
The operators $\mathcal{O}$  are dual to a set of bulk fields $V_{bulk}$ and the boundary value evaluated at $z=L_0$ of the bulk fields are the sources $V_0$.

 The currents that encode supersymmetry breaking will couple to the broken CFT states and not to the source fields directly:
\be
\mathcal{L}\supset \int d^5 x   \sqrt{-g} \ g_{IR} \int d^4 \vartheta  J_{\cancel{SUSY}}V_{bulk} \delta(z=IR). \label{3}
\ee
where one expects the spontaneous breaking of supersymmetry to be an IR effect parameterised by a set of currents  $J_{\cancel{SUSY}}$. The coupling $g_{IR}$ has been introduced which can be defined later. One may either work with current correlators that simply parameterise the supersymmetry breaking effects on the IR brane, or introduce new degrees of freedom which will play the role of a Goldstino multiplet coupled to some messenger fields.   Indeed the resulting effective action arising from ``figuratively'' taking correlators of \refe{3}  should encode also breaking supersymmetry by the geometry. If one takes the messengers literally, they are highly composite degrees of freedom in the IR of the theory. These will be represented by chiral superfields localised in the IR.   One may of course relax the assumption of a delta function in \refe{3} at the cost of introducing new degrees of freedom to the bulk action.
 Similar constructions have been explored in flat five dimensional models and quiver models \cite{McGarrie:2010kh,McGarrie:2010qr,McGarrie:2011av,McGarrie:2011dc}. 

There are some particular criteria for the supersymmetry breaking sector and in particular the messenger fields. We are interested only in models where supersymmetry is broken spontaneously.  Firstly one should be able to identify a Goldstino mode such that supersymmetry is truly spontaneously broken. Secondly as one is working in the limit that supergravity corrections are switched off it is expected that the messenger sector satisfies $\text{Str}\mathcal{M}^2=0$.  This model satisfies these criteria by construction.  The scale of supersymmetry breaking $\hat{M}$ will be associated to a pseudo modulus as the superpartner of a Goldstino.  As in dynamical supersymmetry breaking models this vev and the scale of the magnetic gauge fields, given by $m_{kk}\sim(n-1/4 )\pi/L_1$, are of the same order such that $m_{kk}/\hat{M} \sim O(1)$.

The structure of this paper is as follows: In section \ref{ACTION} the super Yang-Mills bulk action on a slice of $AdS_5$ is outlined and the parameters and operators of the weakly gauged flavour symmetry of a large $N_c$ approximate CFT are matched.  In section \ref{EQUATIONS}  the equations of motion of the free bulk fields are used to find their holographic decomposition.  In section \ref{BOUNDARY} the  boundary action is determined and used to compute the \emph{supersymmetric}  effective action from two point functions of boundary correlators $\braket{O(x)O(0)}$.  Through inversion, these will determine the Green's functions of the free fields and allow for the computation of  the beta function of the gauge theory from the boundary action.   In section \ref{BROKEN} we encode a supersymmetry breaking sector on an IR brane encoded in bulk currents and compute soft masses.    In section \ref{SCATTERING} cross sections and form factors are explored.  In section \ref{CONCLUDE}  we conclude and discuss possible extensions.  
In appendix \ref{BULKFIELDS} the response of a bulk and boundary field to different types of sources is reviewed.  This will determine  the bulk Green's functions. Appendix \ref{APP1}  outlines the notations and conventions used.

\section{The action}\label{ACTION}
In this section we construct a holographic model on a slice of $AdS_5$. This setup  is meant as an effective description of some, as yet unknown, strongly coupled large $N_c$ approximate CFT \cite{ArkaniHamed:2000ds,Rattazzi:2000hs,Gherghetta:2010cj,Gherghetta:2006ha}. The conformally flat metric of this model is given by
\be
ds^2=a^2(z) (\eta^{\mu\nu}dx_{\mu}dx_{\nu}+dz^2)\label{metric}
\ee
where the Minkowski metric is mostly positive $\eta^{\mu\nu}=\text{diag}(-1,1,1,1)$
and 
\be
a^2(z)=\left( \frac{R}{z}\right)^2.\label{aa}
\ee
Setting $R=1$ one obtains $a(z)=1/z^2$. The fifth coordinate $z$ has range 
\be
 L_0<z<L_1,
\ee
where the direction $z$ corresponds to the energy scale, such that $z=L_0$ is the UV AdS boundary and $z=L_1$ the IR brane.  It is useful to define
\be
 G_{MN}=a^2(z)\eta_{MN} , \ \ \ G^{MN}=a^{-2}(z)\eta^{MN}, \ \sqrt{-g}= a(z)^5.
\ee
To define spinors  in this space it is useful to identify the f\"unfbein components
\be
e^a_{\mu}= a(z) \delta^a_{\mu} \ \  e^{\mu}_a= a^{-1} \delta^{\mu}_a \  \  \    e^5_{\mu}=0  \ \ \  e^{\hat{5}}_5=a(z)\delta^{\hat{5}}_5
\ee
where $G^{MN}=e^M_A e^N_B\eta^{AB}$.  It is natural to also define theta warped coordinates \cite{Bagger:2011na,McGarrie:2010yk}
\be
\vartheta= \theta a^{1/2}(z)   \ \  \text{and}  \ \ \int \! d^2\vartheta=a^{-1}(z)\int\!  d^2\theta .
\ee

Following the holographic correspondence one expects that a  flavour symmetry of the approximate CFT corresponds to a gauge symmetry in the bulk AdS space. To weakly gauge the flavour symmetry one positions a UV brane a small distance $z=L_0$ away from $z=0$ which results in a normalisable massless mode.
We wish to explore the flavour symmetry in the AdS background and assume that supergravity corrections are small.  The five dimensional $\mathcal{N}=1$ super Yang Mills action is given by \cite{Mirabelli:1997aj,Hebecker:2001ke,Marti:2001iw,Cacciapaglia:2008bi}
\be
S_{5d}=\int d^5x \sqrt{-g}\frac{1}{g^2_{5d}}\mathcal{L}
\ee
with
\be
\mathcal{L}=-\frac{1}{4} F_{MN}F^{MN}-\frac{1}{2}D_M\Sigma D^M\Sigma-\frac{i}{2}(\bar{\lambda}^i\gamma^MD_{M} \lambda_i- D_{M} \bar{\lambda}^i\gamma^M\lambda_i ) \nonumber
\ee
\be+ m_{\lambda}\bar{\lambda}^i\lambda_i + \frac{1}{2}m^2_{\Sigma}\Sigma \Sigma+(X^a)^2+ \bar{\lambda}_i[\Sigma,\lambda^i]\label{thesusyaction}
\ee
where a $\text{Tr}$ over gauge indices is implicit.  The field content is a positive parity vector multiplet V and a negative parity adjoint chiral multiplet $\Phi_{adj}$.  In five dimensions, $R^{1,4}$, the theory has 8 supercharges, or $\mathcal{N}=2$ from the four dimensional perspective. After implementing the $AdS_5$ metric, the fixed points at $z=L_0$ and $z=L_1$ result in a theory that preserves only $4$ supercharges, or $\mathcal{N}=1$ in four dimensions on an $R^{1,3}\times S^1/\mathbb{Z}_{2}$ background.


The fermions are symplectic Majorana with an $SU(2)_R$ index as defined in \cite{Hebecker:2001ke} such that
\begin{equation}
\lambda^1=\left(
  \begin{array}{c}
   \lambda_{L,\alpha}\\
\bar{\chi}^{\dot{\alpha}}_{R}
  \end{array}
\right),\qquad
\lambda^2=\left(
  \begin{array}{c}
    \chi_{R\alpha}\\
-\bar{\lambda}^{\dot{\alpha}}_L
  \end{array}
\right).
\end{equation}
The fermion $\lambda_L$ is of positive parity the fermion $\chi_R$ is negative and therefore no Dirac mass for the type  $m_D\lambda^0\chi^0$ may arise.  There will be mixing of $\lambda^0$ with the CFT states $\chi^n$ however \cite{Marti:2001iw,ArkaniHamed:2001mi}.

\subsection{Parameters and scales}
It is useful to discuss the relevant parameters and scales in relation to holography and a string theoretic embedding of this construction.  Let us  assume the four dimensional approximate CFT is a  large $N_c$ gauge theory with gauge coupling $g_{YM}$.  The 't Hooft coupling is given by $\lambda_t= g^2_{YM} N_c$.  These scales may be related to the string scales:  the string coupling  is given by $g_s=g^2_{YM}$ and the string length $l^2_s=\alpha'$.  This allows one to set the AdS curvature radius $R^4= 4\pi l^4_s g_s N_c$ \cite{Maldacena:1997re,Witten:1998qj,Aharony:1999ti}.

The strong coupling limit of the four dimensional gauge theory is where $\lambda_t \gg 1$.  Taking the large $N_c$ limit $g_{YM}\rightarrow 0$.  In terms of the string scale $g_s=g^2_{YM}\rightarrow 0$ in which case one finds the supergravity limit that 
\be
\left(\frac{R}{l_s}\right)^4 =4\pi  g_s N_c = 4\pi \lambda_t \gg 1.
\ee
In this limit one finds a weakly coupled effective supergravity description of the strongly coupled gauge theory, in a warped higher dimensional space.

There are also two other parameters which need to be discussed. The first is $g_5$. As the five dimensional gauge theory is living in the supergravity background, it should be possible to relate it to the fundamental scales of the four dimensional gauge theory.  This is achieved by matching, at tree level,
\be
\frac{g^2_5}{R}=\frac{12\pi^2} {N_c}\sim 1/F_{n}^2.
\ee
The decay constants for the nth vector meson is labelled $F_n$ appearing in \refe{OO}.  The boundary effective action that we compute in section \ref{BOUNDARY} are tree level effects in the large $N_c$ limit.  We  also comment on the parameter $g_{SM}$ corresponding simultaneously to the coupling of the standard model groups $U(1)\times SU(2)\times SU(3)$, where we are for the most part effectively computing for the $U(1)_{em}$ case.  This is an external coupling $g_{SM}$  associated to the external sources not related to the $AdS/\cancel{SUSY}$ construction and we may work perturbatively in  $\alpha_{SM}$.  In particular, the supersymmetric standard model and the $AdS/\cancel{SUSY}$ setup completely decouple in the limit  $\alpha_{SM}\rightarrow 0$.  Typically the one loop beta function can relate $g^2_5/R \sim g^2_{SM}\log (L_1/L_0)$.  For a Planck scale to electroweak scale hierarchy $\sim 10^{16}$ one finds  $N_c$ between 1 and 10.   For an $M_{Planck}$ to $M_{SUSY}$ hierarchy,  $N_c$ can be larger at the cost of lifting $M_{SUSY}$.  We expect that $m_{kk} \sim 1/L_1$ and $M_{SUSY}$ are of the same order.

In general it is possible to compute n point functions, for example 
\be
\braket{\mathcal{O}\cdot\cdot\cdot\mathcal{O}}=\frac{\delta^n\text{Ln}Z[A_0]}{\delta A_0\cdot\cdot\cdot \delta A_0}
\ee
where $A_0^{\mu}$ acts as a source for the operator $\mathcal{O}^{\mu}$ analogous to $\bar{q}\gamma^{\mu}q$.  In terms of Witten diagrams these n point functions are represented as contact diagrams.  There are also exchange diagrams.  Some parameters are fixed by the above.  Taking the two point function of the boundary to boundary correlator one obtains
\be
\braket{\mathcal{O}(p)\mathcal{O}(-p)}=\sum_n \frac{F_n^2}{p^2+m_n^2+i\epsilon} \label{OO}.
\ee
As reviewed in \cite{Witten:1979kh}  the current correlator \refe{OO} contains only colourless one meson states in the planar limit.  This will give a decay constant $F_n=\braket{0|\mathcal{O}|n}$ and also a mass $m_n$ to create an nth meson from the vacuum.  It is expected that $F_n\sim \sqrt{N_c}\sim \sqrt{ L/g_5}$ so there should be no issue with the perturbative expansion in this way \cite{Hooft1974461,Witten:1979kh}. In terms of embedding into string theory, the vector mesons should be associated with the open string sector as opposed to the glueball spectrum of the closed string sector.

\subsection{Identifying operators}
Let us now discuss the identification of various operators.  In the large $N_c$ limit there are in principle an infinite number of operators in the 4d strongly coupled gauge theory.  
In the limit $z\rightarrow 0$ one is able to identify the boundary value of a bulk field with conformal dimension  $\Delta$ of the $p$ form of dimension $d$
\be
\lim_{z\rightarrow L_0 }\left( z^{\Delta -d+p}X(x^{\mu},z)\right)\rightarrow   X^0(x^{\mu}) \label{bbcorrespondence}
\ee
which acts as a source for the bulk field and therefore treats the bulk field as a current $X(x,z)$ coupled to a source which are then both treated independently.  The four dimensional operators have scaling dimension $\Delta$ and the bulk p-form field  such that the AdS mass $m^2$ is determined by \cite{Gubser:1998bc,Witten:1998qj} 
\be
R^2 m_{AdS}^2= (\Delta-p)(\Delta+p-4)
\ee
and for the fermions
\be
R^2 m_{AdS}^2=\Delta(\Delta-d).
\ee
We therefore wish to introduce coupling between source fields and bulk fields
\be
g_{SM} \int d^4 \theta \mathcal{O}V_{0}\label{coupling}
\ee
where $g_{SM}$ is a coupling unrelated to the $AdS$ parameters and where one can work perturbatively in $\alpha_{SM}\rightarrow 0 $, the decoupling limit of GGM \cite{Meade:2008wd}.  As in AdS/QCD models, we wish to match the UV (S)QCD flavour currents to bulk gauge fields, which then couple to the external sources.   These currents are made of the UV squark and quark fields. In this paper we wish to take the diagonal flavour group, essentially $\mathcal{O}^{\mu}=\frac{1}{2}\left(\mathcal{O}^{\mu}_{L}+\mathcal{O}^{\mu}_{R}\right)$, such that  it is evident that $A_{\mu}$ is the vectorial or diagonal of a possible $SU(N_f)_L\times SU(N_f)_R$ flavour symmetry, which will each have their own dual field $A^{\mu}_{L,R}$.    This allows one to match \cite{Maldacena:1997re,Witten:1998qj,Gubser:1998bc,Aharony:1999ti,Erlich:2005qh} a set of operators to a set of bulk fields as in table \ref{table}.

\begin{table}
\begin{center}
\begin{tabular}{|c c c | c|c|}
\hline
4D:  operator & & Field& $\Delta$ &$m^2$ \\
\hline
$\mathcal{O}_{L,R}=\phi^{\dagger}\phi_{L,R} $&$\rightarrow $&$ D(z,x) _{L,R} $ &2  &-4\\
$\mathcal{O}^{\alpha}(x)_{L,R} =-i\sqrt{2}\phi^{\dagger}q^{\alpha}_{L,R} $&$\rightarrow $&$  \lambda^{\alpha}(z,x)_{L,R}    $& 5/2&1/2\\
$\mathcal{O}^{\mu}(x)_{L,R}=\bar{q}\sigma^{\mu}q_{L,R}-i\left(\phi^{\dagger}\partial^{\mu}\phi-\partial^{\mu}\phi^{\dagger}\phi \right)_{L,R} $&$\rightarrow $&$A^{\mu}(z,x) _{L,R}   $&3&0\\
\hline
\end{tabular}
\caption{Operators corresponding to the bulk fields of the model.}\label{table}
\end{center}
\end{table}
Whilst the bulk boundary correspondence \refe{bbcorrespondence} holds for fields with positive parity on the boundary, there is no equivalent correspondence for the negative parity fields such as the degrees of freedom in\footnote{There may be non dynamical sources.}  $\Phi_{adj}$.  These fields do not have (broken) CFT operators \cite{Cacciapaglia:2008bi} and are emergent degrees of freedom.  There may be negative parity bulk fields or IR brane localised fields as well. Conversely if one switches boundary conditions there are no sources for $V$ \cite{Cacciapaglia:2008bi}.  It is useful to highlight from the outset that the above currents in the table are not the currents associated with supersymmetry breaking currents. There are also models \cite{McGuirk:2011yg,Skenderis:2012bs,Argurio:2012cd,Argurio:2012bi} in which the geometry breaks supersymmetry, but where these fields are identified \emph{directly} with the supersymmetry breaking currents. We expect that our prescription, of an IR or bulk effective action, should apply also to those papers, even if supersymmetry is broken by the geometry.   It is satisfying to see that this is also confirmed by holographic deconstructions \cite{Son:2003et,Csaki:2001em,McGarrie:2010qr}.

The supersymmetry breaking currents will be labelled  $\mathcal{J}$ as originally in \cite{McGarrie:2010yk,McGarrie:2011av,McGarrie:2012ks}.   To motivate these currents $\mathcal{J}$, let us compare to the case of AdS/QCD \cite{Son:2003et,DaRold:2005ju,Erlich:2005qh,Karch:2006pv}.  In those models the boundary  operator $\bar{q}q$ is mapped to the pion bulk field $X(x,z)$, which then supplies a bulk pion current $J_{\pi}=\pi\partial_{\mu}\pi$. This construction naturally allows for the computation of cross section $\sigma(e^+,e^-\rightarrow \pi^+,\pi^-)$. We expect that the SUSY breaking currents arise in a similar fashion, as is also discussed in \cite{Abel:2010vb}, in which the $\mathcal{J}$ currents are extracted from a ``magnetic'' description on the IR brane.

The currents $\mathcal{J}$ are able to encode the spontaneous breaking of supersymmetry.    In particular these currents ($\mathcal{J}$) do not directly couple to the external sources but to the approximate CFT states, which is important for a phenomenologically viable model and to encode the expected form factors of vector mesons.  

The final, ``all orders" current correlators that couple to external sources may be found from taking 
\be
\frac{d^2S_{eff}[V_0]}{dA^{\mu}_0(x) dA^{\nu}_0(y)}=\braket{K_{\mu}(x)K_{\nu}(y)}_{full}\label{full5}
\ee
after fully integrating out the bulk and IR boundary actions. In analogy to QCD, this correlator may be compared with the total hadronic vacuum polarisation amplitude $\Pi_{\text{had}}(s)$. This correlator receives pieces from \refe{effectiveaction2} and \refe{effectiveaction3} (and higher order contributions).  The parts in \refe{effectiveaction3} contribute the predominant effect of the running gauge coupling  and the parts in \refe{effectiveaction3}  are the leading contributions of supersymmetry breaking effects after integrating out the IR brane.
\section{The fields}\label{EQUATIONS}
In this section we wish to take the free field equations of motion for the bulk theory and determine the holographic decomposition.  This will allow one to compute boundary correlators for the supersymmetric theory and then when we introduce supersymmetry breaking sources, will allow one to carry out perturbation theory to compute higher order correlators.   Some of the results in this section have been collated from work on holographic QCD. The starting point is to take the free bulk equations of motion.

\subsection{Equations of motion}
The vector superfield of the bulk gauge theory contains  $A_{\mu}$, $\lambda_{\alpha} $ and $D=(d_{z}\Sigma-X^3)$ as dynamical degrees of freedom.  As the bulk theory is $\mathcal{N}=1$ in five dimensions there is also an adjoint chiral superfield $\Phi_{adj}$ which contains $(A_5+i\Sigma)$ and $\chi$ as dynamical degrees of freedom.  As $\lambda$ and $\chi$ are related as a symplectic Majorana fermion, determining $\lambda$ completely fixes the properties of $\chi$.

 The equations of motion for the gauge field is given by \cite{Gherghetta:2000qt}
\be
\partial_M(\sqrt{-g}g^{MN}g^{PQ}F_{NQ}(x,z))=0\label{gaugeeom}
\ee
taking $A_z=0$ and working in $\partial_{\mu}A^{\mu}=0$ gauge. For the symplectic Majorana spinor  the equation is 
\be
\left( g^{MN} \gamma_M D_N+m_{\Psi } \right) \Psi (x,z)=0
\ee
A real scalar field $\Sigma$ or $A_5$ may be modelled by 
\be
\frac{1}{\sqrt{-g}}\partial_M\left(\sqrt{-g}g^{MN}\partial_N \Sigma(x,z)\right)-m_\Sigma^2 \Sigma(x,z)=0\label{eom1}
\ee
where it is possible to relate this to the  dynamical part of the  D-term $D=d_{z}\Sigma$ as pointed out in \cite{McGarrie:2010kh}.

Using the metric \refe{metric}   gives the equations of motion
\be
\left[\eta^{\nu\rho}\partial_{\nu}\partial_{\rho}+a(z)^{-s}\partial_z a^{s}(z)\partial_z- a^2M_{\Phi}^2\right]\Phi(x,z)=0 \label{yumyum}
\ee
where $\Phi= \{A_{\mu},\Sigma \}$ with $s=\{1,3\}$  $M_{\Phi}^2=\{0,m^2_{\Sigma} \}$. For the fermions one finds a coupled equation
\be
\left[\partial_z +2a^{-1}(z)\partial_z a(z)\pm a(z)m_{\Psi} \right]W_+/W_-(x,z)=\pm p \ W_-/W_+(x,z)\label{eom2}
\ee
where $ \{\lambda,\chi \}$ is related to $\{W_+,W_-\}$ respectively.  This is solved by finding the appropriate independent second order equations \cite{Contino:2004vy}.

In the Kaluza-Klein decomposition one finds solutions by identifying $\partial^2 f_n(z)=-m_n^2f_n(z)$.  In the holographic basis one leaves the $p^2$ dependence.  Let us discuss the KK spectrum of this theory:  there are massless modes for $A_{\mu}$ and $\lambda$, as well as a tower of massive modes. Once one introduces the supersymmetric standard model on the UV boundary (outside the AdS system) the massless modes will correspond to standard model gauge fields. There are no massless modes for $\chi$, $\Sigma$ and $A_5$.  
The massive modes arise as part of a supersymmetric Higgs mechanism including Dirac masses between $\chi$ and $\lambda$. The massless modes are related to the external fields of the theory which leads us to discuss the holographic decomposition.

\subsection{The holographic decomposition}
We wish to specify the holographic decomposition of the bulk fields. To do this one first identifies the boundary fields $A^0, \lambda^0, D^0=d_z\Sigma$,  which couple to the four dimensional operators $\mathcal{O}_{\mu},\mathcal{O}_{\alpha},\mathcal{O}$ respectively and which correspond to the bulk fields $A_{\mu}(x,z), \lambda_{\alpha}(x,z), d_{z}\Sigma(x,z)$, as in table \ref{table}.  The bulk fields are determined in terms of Bessel functions $\{J_{\alpha},Y_{\alpha}\}$ of the first and second kind.

\subsubsection{The gauge field}
For the gauge field to have a massless zero mode it should have Neumann boundary conditions on \emph{both} branes.  This in turn fixes constraints on the fields in the same vector multiplet.   Additionally this sets $A_5$ to have Dirichlet boundary conditions to cancel boundary terms, fixing the conditions of $\Phi_{adj}$ in the process. The function $V(q,z)$ may be related to Migdal's Pade approximation of the OPE \cite{Erlich:2006hq}.   Taking Neumann boundary conditions for the vector $\partial_z V=0$ we obtain  for the gauge field \cite{Erlich:2006hq,Grigoryan:2007vg}
\be
A_{\mu}(q,z)=A^{0}_{\mu}(q)\frac{V(q,z)}{V(q,L_0)}=A^{0}_{\mu}(q)K(q,z)
\ee
with 
$A(p,L_0)=A^0(p)$ implemented and in addition
\be
A_{\mu}(q,z)=\int d^4 x e^{iq.x}A_{\mu}(x,z).
\ee
The bulk to boundary propagator or profile function is $V(q,z)$, as solution of \refe{yumyum}  with 
\be
V(q,z)= z q\left[Y_0(q L_1) J_1(q z)-J_0(q L_1)Y_1(q z) \right] \label{bb}
\ee
where $q=\sqrt{q^2}$ and $V_{0}^{\mu}=V_{0} \epsilon^{\mu}(q)$ and $V^{\mu}(x,z)=\int d^4 q e^{-iq.x} \  V^{\mu}(q,z)$. The derivative gives an odd function \cite{Erlich:2006hq}
\be
\partial_z V(q,z)=z q^2\left[Y_0(q L_1) J_0(q z)-J_0(q L_1)Y_0(q z) \right].
\ee
The same construction may be applied to the fermions.
\subsubsection{The fermions}
The fermion may be split into two 2 component spinors, one of even and one of odd parity.   Taking the ansatz for \refe{eom2}
\be
\lambda_{\alpha}(q,z)=\frac{1}{W_{+}(q,L_0)}\lambda^0_{\alpha}(q)W_{+}(q,z)
\ee
\be
\chi_{\alpha}(q,z)=\frac{1}{W_{-}(q,L_0)}\chi^0_{\alpha}(q)W_{-}(q,z)
\ee
and substituting into the equations of motion
\be
\gamma^{\mu}q_{\mu} W_{+}(q,z) - \partial_zW_{-} + (c+2)W_{-}(q,z)=0
\ee
Due to the above equations it is straightforward to see that the boundary conditions 
\be
\delta \bar{\lambda} \chi |_{0,\pi R}=\delta \bar{\chi}\lambda |_{0,\pi R}=0
\ee
require that one cannot have a massless mode for both fields. Therefore a Dirac mass for the erstwhile massless  modes cannot arise.

The Dirac equation is satisfied by relating the source $\lambda^0$ to $\chi^0$
\be
\sigma^{\mu}_{\alpha\dot{\alpha}}p_{\mu}\bar{\chi}^0=p \frac{W_{-}(q,L_0)}{W_{+}(q,L_0)}\lambda^0_{\alpha} \label{dirac1}
\ee
\be
\sigma^{\mu}_{\alpha\dot{\alpha}}p_{\mu}\bar{\lambda}^0=p \frac{W_{+}(q,L_0)}{W_{-}(q,L_0)}\chi^0_{\alpha}.
\ee
The solutions  are \cite{Contino:2004vy} 
\be
W_{-}(q,z)= z^{5/2}  \left[ J_{\alpha-1}(q z) Y_{\beta}(q L_1)- J_{\beta}(q L_1)Y_{\alpha-1}(p z)\right]
\ee
for the odd case
\be
W_{+}(q,z)=z^{5/2}   \left[ J_{\alpha}(q z) Y_{\beta}(q L_1)- J_{\beta}(q L_1)Y_{\alpha}(p z)\right]
\ee
for the even case, where 
\be
\alpha=m_{\psi}R+ 1/2=c+1/2
\ee
\be
\beta = \alpha-1.
\ee
In fact we shall see that the conditions of positive parity for the multiplet $V$ and supersymmetry uniquely fixes $\beta=0$ and $\alpha=1$. Independently we expect $\Delta=5/2$ for the operator field correspondence where $\Delta=3/2+|c+1/2|$  which also predicts $c=1/2$, where $c=m_{\Psi}R$.  Let us now discuss the scalar fields.

\subsubsection{The scalar}
The negative parity $\Sigma$ scalar couples to the boundaries through $\partial_z\Sigma$.  One identifies a scalar through $\partial_z\Sigma(q,z)=D(q,z)$
\be
\partial_z\Sigma(q,z)= D^0(q)\frac{E(q,z)}{E(q,L_0)}
\ee
where  $D(q,L_0)=D^0(q)$.   It is required that the super-traced combination of current correlators of the effective action in \refe{supertrace} to vanish.  This requirement fixes the form of the fields up to an overall normalisation.  This gives an odd function
\be
E(q,z)=N(z) \left[ Y_{0}(q L_1)J_{\tau}(q z)-J_{0}(qL_1)Y_{\tau}(q z)\right]
\ee
A condition is $\text{Dim}[\mathcal{O}]=2$, which relates to $\partial_z \Sigma$  where $\text{Dim}[\mathcal{O}]=2+\tau$, setting $\tau=0$ for a general scalar  explored in \cite{DaRold:2005vr,Gherghetta:2006ha}.  This gives an even function
\be
\partial_zE(q,z)=N(z) q\left[Y_{0}(q L_1)J_{1}(q z) - J_0(qL_1)Y_{1}(qz)  \right]
\ee
These field profiles will now be used to evaluate the boundary action.


\section{The boundary action}\label{BOUNDARY}
\begin{figure}[ht]
\begin{center}
\includegraphics[scale=0.8]{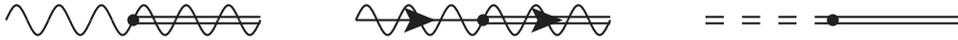}
\caption{A diagram to represent the external field couplings to bulk vector mesons in \refe{currentsource}, on the UV boundary.}
\label{currenttosources}
\end{center}
\end{figure}
In this section the procedure for determining the boundary action is reviewed.  As already stated one determines the boundary value of the bulk fields to be sources of the bulk fields which one then treat as independent. This allows one to identify
\be
g_{SM}\int d^4 \theta  V^0 \mathcal{O}(x) \label{currentsource}
\ee
where $V^0(x)$ is the boundary gauge theory and $\mathcal{O}(x)$ is a current multiplet.  This is represented in figure \ref{currenttosources}.   One should therefore determine the boundary action for the gauge fields \cite{Gherghetta:2006ha,Batell:2007ez,Batell:2007jv} and for fermions \cite{Arutyunov:1998ve,Henneaux:1998ch,Contino:2004vy}. The boundary conditions for boundary actions have also been explored in \cite{Belyaev:2005rs,Belyaev:2005rt}.

Labelling the source fields
\be
A_{\mu}^0(x)=A_{\mu}(x,z=L_0)  \  , \  \lambda^0(x)=\lambda(x,z=L_0) \  , \  D^0(x) = \partial_z \Sigma(x,z=L_0).
\ee
One cannot also fix  a true source for $\chi(x,z)$, $\chi^0$,  on the UV boundary as the fermions of a symplectic Majorana spinor are not independent \cite{Contino:2004vy} and $\chi$ vanishes on the UV boundary.  This is our first example of  bulk fields without a proper boundary source.  

In addition the variation of the bulk action generates a boundary action which is required to vanish.  This sets $\chi(L_1)=0$ and  sets Neumann boundary conditions on the IR brane
\be
\partial_zA_{\mu}(x,L_1)=0  \  , \  \partial_z\lambda(x,L_1)=0 \   ,  \  \partial_zD(x,L_1) = 0.
\ee
With the above identification our procedure is as follows:  

\begin{itemize}
\item  Set the variation of the five dimensional action, with respect to every field, to vanish: $\delta S_{5d} + \delta S_{boundary}=0$.
\item  In addition implement Neumann boundary conditions for the positive parity multiplet $V$ and Dirichlet for $\Phi_{adj}$. This  fixes the freedom of the fields at the UV and IR.
\item Use integration by parts, for which the equations of motion vanish leaving only the boundary term.
\item The resulting action  should still be supersymmetric under $\mathcal{N}=1$ four dimensional supersymmetry.
\end{itemize}

It is useful to compare this approach with the Kaluza-Klein approach.  As commented before \cite{Contino:2004vy}   the cancellation of $\delta S_{5d}=0$ is automatic if the Dirichlet condition of even parity ($+$) is used for $\lambda$, exactly in correspondence with the warped models \cite{McGarrie:2010yk,McGarrie:2012ks}.  

Following the above procedure, the resulting action is \cite{Belyaev:2005rs,Belyaev:2005rt,McGuirk:2011yg}
\be
\frac{1}{g_5^2}\! \int \! d^4x \sqrt{-g_{\text{5d}}} \left( - \frac{1}{4}G^{zM}G^{PQ}F_{MQ}A_{P}  +\frac{i e^{\hat{z}}_z} {2}\bar{\lambda}^i \lambda_i -\frac{1}{2}G^{zM}(\partial_M \Sigma (x,z)) \Sigma(x,z)  \right)|^{z=L_1}_{z=L_0} \label{eff}
\ee
where $e^{\hat{z}}_z=a(L_0)\delta^{\hat{z}}_z$. This gives for the UV boundary
\be
 \frac{1}{g_5^2}\int \! d^4x \! \left( \frac{a(z)}{2} (\eta^{\mu\nu}A_{\mu}\partial_z A_{\nu}-2\eta^{\mu\nu}A_{\mu}\partial_{\nu}A_5)  \! +\! \frac{ia^4(z) } {2}\bar{\lambda}^i \lambda_i +{a^3(z)}(\partial_z \Sigma (x,z)) \Sigma(x,z)  \right)_{z=L_0}\label{eff2}
\ee
 There are a few additional operators  which may be added  \cite{DaRold:2005ju,Cacciapaglia:2008bi}, but this is a minimal choice 
which can be related to the Gibbons-Hawking-York boundary term \cite{Belyaev:2005rs,Belyaev:2005rt} and these terms are introduced precisely to make the bulk action supersymmetric in the presence of boundaries. Taking a Fourier transform one can equivalently define
\be
\frac{1}{g_5^2}\int \! \frac{d^4p}{(2\pi)^4} \left[\! \frac{a(z)}{2} (\eta^{\mu\nu}A_{\mu}(p,z)\partial_z A_{\nu}(-p,z)-2\eta^{\mu\nu}A_{\mu}(p,z)\partial_{\nu}A_5(-p,z))  \right] |_{z=L_0}
\label{eff4}
\ee
\be
 + \frac{1}{g_5^2}\int \! \frac{d^4p}{(2\pi)^4} \left[\! \frac{ia^4(z) } {2}\bar{\lambda}^i(p,z) \lambda_i (-p,z)+{a^3(z)}(\partial_z \Sigma (p,z)) \Sigma(-p,z)\right] |_{z=L_0}.\nonumber
\ee

For a small finite $L_0$ (instead of taking the UV cutoff to infinity) the  source fields are normalisable and a kinetic term for these are generated \cite{ArkaniHamed:2000ds}. This gauges the erstwhile global symmetry on the boundary.  In expectation of this one may introduce a UV action for the sources
\be
S_{UV}\supset \int d^4 x   \  \sqrt{-g_{\text{ind}}}  \left(  \  \bar{\lambda}'_0\bar{\sigma}^{a}e_a^{\mu}\partial_{\mu}\lambda'_0   +G^{\rho\nu} G^{\sigma\nu}F_{\rho \sigma}F_{\mu\nu} + a^{-2}(L_0)(D'_0)^2  \right  )
\ee
where $\sqrt{-g_{\text{ind}}}=a(L_0)^4$ , related to the induced metric on the UV boundary.  The last one is inferred from a typical scalar
\be
\int d^4x \sqrt{-g_{\text{ind}}}\left( G^{\mu\nu}D_{\mu}\phi D_{\nu}\phi^*\right).
\ee
The sources should be normalised so that the boundary kinetic terms will be canonical, giving 
\be
\lambda_0= \lambda'_0 a^{3/2}(L_0) \ \  , \ \ D_0= D'_0a(L_0)    \  , \ \ A_0= A_0   .  \label{rescale}  
\ee 
It is now possible to compute the tree level effective action.
\subsection{The supersymmetric effective action}
Now that we have a boundary action and implemented the holographic principle it is possible to compute the current correlators of the supersymmetric effective action.  These are pictured in figure \ref{wittendiag0}.  Individually these correlators have been explored before in \cite{ArkaniHamed:2000ds,Agashe:2004rs,Contino:2004vy,DaRold:2005ju,Gherghetta:2006ha}.  The poles of the correlators correspond to the mass spectrum.    The inverse of these correlators correspond to a UV boundary to UV boundary Green's functions.  It is useful to stress that these two point functions do not encode supersymmetry breaking. Of course once supersymmetry is broken spontaneously it will shift the poles of the bulk fields.  This will shift the poles in these correlators, which will appear as an \emph{explicit} breaking of supersymmetry, and a \emph{subleading} effect of this breaking will appear in these correlators: one should keep track of the factors of $g_5$.

\begin{figure}[ht]
\begin{center}
\includegraphics[scale=0.5]{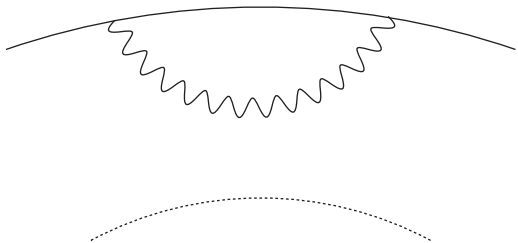}
\caption{A Witten diagram for a current correlator  $\braket{\mathcal{O}_{\mu}\mathcal{O}_{\nu}}$.  Operators $\mathcal{O}_{\mu}$ on the UV boundary (solid line) correspond to a gauge field $A_{\mu}$ in the bulk.  The dotted line is the IR boundary.  This contribution is $O(N_c)$.
} \label{wittendiag0}
\end{center}
\end{figure}
To compute the boundary correlators we should compute for instance
\be
\frac{\delta^2 S_{5d}[V^0(x)]}{\delta V^0_{\mu}(x) \delta V^0_{\nu}(0)}=\braket{O_{\mu}(x)O_{\nu}(0)} \label{correl}
\ee
where the Fourier transformed vector current correlator is given by 
\be
\int d^4 x e^{iq.x}\braket{O_{\mu}(x)O_{\nu}(0)}=(q^2 g_{\mu\nu}-q_{\mu}q_{\nu})\Pi_1(q^2)
\ee
\be
\int d^4 x e^{iq.x}\braket{O_{\alpha}(x)O_{\dot{\alpha}}(0)}= (\sigma^{\mu}_{\alpha\dot{\alpha}}\partial_{\mu})\Pi_{1/2}(q^2).
\ee
The boundary correlators of \refe{correl} may be related by
\be
\braket{O(q)O(-q)}_a=\lim_{L_0\rightarrow 0 } \left(q^2 \Pi(q^2)_a +\text{UV counter terms}\right)
\ee
with $a=0,1/2,1$. The UV counterterms cancel any local divergent terms in the limit $L_0\rightarrow0$ \cite{Contino:2004vy}. 
In momentum space this will generate a supersymmetric effective action on the UV boundary
\be
\int \frac{d^4q}{(2\pi)^4}\left[ \frac{1}{4}\Pi_1(q^2)F_{\mu\nu,0} F^{\mu\nu}_0-i\Pi_{1/2}(q^2)\lambda_0\sigma_{\mu}\partial^{\mu}\bar{\lambda}_0
+ \frac{1}{2} \Pi_0(q)D_0^2
\right]
\ee
where for instance $D^2_0$ is understood to mean $D_0(q)D_0(-q)$. In position space this may be rewritten
\be
\delta L^{SUSY}_{eff}= \frac{g^2_{SM}}{2} \Pi_0(0)D^2_0 -ig^2_{SM} \Pi_{1/2}(0)\lambda_0\sigma_{\mu}\partial^{\mu}\bar{\lambda}_0-\frac{g^2_{SM}}{4} \Pi_{1}(0) F_{\mu\nu,0}F^{\mu\nu}_0.  \label{effectiveaction2}
\ee
A rescaling  $V^0\rightarrow g_{SM}V^0$ has been made to obtain this expression.  This action is at order $O(1/g_5^2)$ or equivalently $O(N_c)$. This is the first set of correlators which we will discuss. The scalar correlator is given by 
\be
\Pi^D_0(q^2)= \frac{1}{g_5^2}  \left( a^{3-2}(z)  \frac{E(q,z)}{\partial_z E(q,L_0)}   \right)_{z=L_0}\!\!\!\!\!\!\!\!\!\!\! =\frac{a(L_0)}{qg_5^2}\frac{ \left[Y_0(q L_1) J_{0}(q L_0)-J_0(q L_1)Y_{0}(q L_0) \right]}{ \left[Y_0(q L_1) J_{1}(q L_0)-J_0(q L_1)Y_{1}(q L_0) \right]   } \label{pi0}
\ee 
which couples to $\partial_z \Sigma$ is  given by (after canonically normalising $D_0^2\rightarrow a^{-2}D_0^2$ the boundary action). 
For the gauge fields \cite{Erlich:2006hq,Grigoryan:2007vg}
\be
\Pi^A_1(q^2)= \frac{1}{q^2g_5^2} \left( a(z) \frac{\partial_z V(q,z)}{V(q,L_0)}   \right)_{z=L_0} 
=\frac{a(L_0)}{qg_5^2}\frac{ \left[Y_0(q L_1) J_0(q L_0)-J_0(q L_1)Y_0(q L_0) \right]}{ \left[Y_0(q L_1) J_1(q L_0)-J_0(q L_1)Y_1(q L_0) \right]   }. \label{pi1}
\ee
  For the gaugino  source \cite{Contino:2004vy,DaRold:2005ju,Agashe:2004rs} one finds
\be
 \!\!\!\!\!\!\!\!\!\!   \!\!\!\!\!\!\!\!\!\!  \!\!\!\!\!\!\!\!\!\!  \!\!\!\!\!\!\!\!\!\!   \!\!\!\!\!\!\!\!\!\!  \!\!\!\!\!\!\!\!\!\!  
     \!\!\!\!\!\!\!\!\!\!   \!\!\!\!\!\!\!\!\!\!  
\!\!\!\!\!\!\!\!\!\!   \!\!\!\!\!\!\!\!\!\!  \!\!\!\! \Pi^{\lambda}_{1/2}(q^2)=\frac{1}{qg_5^2} \left(a^{4-3}(z)\frac{ W_{-}(q,z)}{W_{+}(q,L_0)} \right)_{z=L_0}\!\!\!\!\!=\nonumber
\ee
\be
\frac{a(L_0)}{qg_5^2}\left[\frac{J_{\alpha-1}(qL_0) Y_{\beta}(qL_1)- J_{\beta}(qL_1) Y_{\alpha-1}(qL_0) }
{J_{\alpha}(q L_0)Y_{\beta}(q L_1)-J_{\beta}(q L_1) Y_{\alpha}(q L_0)
}\right]\label{fermionone}
\ee
where the $1/g_5^2$ has mass dimension $+1$. The correlators $\Pi_a(q^2)$ are dimensionless. 
As the action is supersymmetric  one expects the supertraced combination to vanish:
\be
\left[3\Pi_1(q^2)-4\Pi_{1/2}(q^2)+ \Pi_0(q^2)\right]\equiv 0 \label{supertrace}
\ee
so comparing \refe{pi1} this sets $\beta=\alpha-1$ and $\alpha=1$  in \refe{fermionone} and with \refe{pi0} fixes $\tau=0$.   Supersymmetry completely determines the boundary conditions and mode expansions of all the fields in the theory and  furthermore it  determines the Green's functions of the boundary theory.    We have reviewed this procedure as we feel it is instructive, however on a more practical level it more appropriate to solve for the Green's functions $\Delta(p^2;z,z')$ and then  one can directly determine the self energies $\Pi(q^2)$ as the inverse of Green's function at $z=z'=L_0$  \cite{Ponton:2012bi}.   It is useful to identify 
\be
\braket{\mathcal{O}(p)\mathcal{O}(-p)} =p^2\Pi(p^2)=\left[ p^2\Delta(p,L_0,L_0) \right]^{-1}.
\ee


There is also in principle the correlator (and its complex conjugate), 
\be
\int d^4x e^{iq.x }\braket{O_{\alpha}(x)O_{\beta}(0)}\equiv0  \label{OOO}
\ee
The exact vanishing of this term is due to the supersymmetry of the action \cite{Argurio:2012cd}.   The five dimensional $\mathcal{N}=1$ action has an $SU(2)_R$ symmetry broken to $U(1)_R$ by  the presence of the branes and resulting boundary conditions on the fermions.   Once the supersymmetry breaking terms are introduced on the IR brane the $U(1)_R$ symmetry is broken in general.  Despite this the correlators \refe{OOO} remains vanishing. One should consider the observed correlator as the sum of the pieces \refe{OOO} and \refe{OO2}.   One could of course harm the bulk and boundary theory to generate such a term, and in so doing harm the closure of supersymmetry transformations of the theory.  A harmful example would be a non canonical kinetic term on the UV boundary
\be
\int d^4 x\sqrt{g_{\text{ind}}}\left( Z' \lambda^{\alpha}\sigma^{a}_{\alpha\dot{\alpha}} e_{a}^{\mu}\partial_{\mu}\bar{\chi}^{\dot{\alpha}}\right)
\ee
(despite $\chi(L_0)=0$ \cite{Contino:2004vy}) then using \refe{dirac1}  and rescaling by \refe{rescale} one finds
\be
\int \frac{d^4p}{(2\pi)^4} \lambda_0^{\alpha}\left( Z' p \frac{W_-(p,L_0)}{W_+(p,L_0)}\right) \lambda_{0\alpha}.
\ee
  As we are not interested in introducing arbitrary explicit breakings of supersymmetry as our goal is to explain the generation of explicit terms in the MSSM from the spontaneous breaking of supersymmetry the above terms are ignored.

If there are massless poles in the above correlators this indicates that the source fields are massless.  The physical spectrum should match the same spectrum as the Kaluza-Klein mode expansion \cite{Contino:2004vy,Gherghetta:2006ha}.  In particular this means there is a massless pole in $\Pi_0$ in \refe{pi0} as the $A^{\mu}_{0}$ source is massless and a massless pole in $\Pi_{1/2}$ in \refe{fermionone} as $\lambda_0$ is massless.  No Dirac mass can arise for $\lambda_0$ as there is no $\chi_0$ mode.
\subsection{Running  gauge coupling}
The above correlators effect the running gauge coupling and change in the beta function and have been explored before \cite{Dienes:1998vg,Pomarol:2000hp,ArkaniHamed:2000ds,Randall:2001gc,Goldberger:2002cz,Goldberger:2002hb,Goldberger:2003mi}.  Taking account of the massless zero modes only, one would expect $b=3N$, where N here is the weakly gauged flavour symmetry $SU(N)$ (not to be confused with the $N_c$ of the broken CFT), which is given by 
\be
b=\frac{11}{3}T(adj)-\frac{2}{3}T(F)-\frac{1}{3}T(S)
\ee
where $T(adj)$ is the index in the adjoint gauge fields, $T(F)$ of the fermions $\lambda_0$ which in this case are also in the adjoint, $T(S)$ for the complex scalar index and a sum over all fields is implicit. Taking into account the running from the full CFT states of $A_{\mu},\lambda_{\alpha}$  \emph{and} one should also take into account the running from $\Phi_{adj}$ which has odd parity.

The broken CFT effects from $A_{\mu}$ may be extracted from $\Pi_1(q^2)$ in \refe{pi1} 
\be
\Pi_1(q^2)= T[R] \frac{R}{2g_5^2}\text{Ln}(q^2 R^2).
\ee
This is determined from the group structure $\braket{O^a O^a}=T[R]\delta^{ab}\Pi$ and the currents are in the adjoint.  The source field gauge coupling of the broken CFT runs logarithmically,  which becomes decoupled by removing the UV brane \cite{ArkaniHamed:2000ds}.  Above the cutoff of this theory one should expect the running from the underlying theory degrees of freedom.
Let us just focus on the running of the external gauge field $A^{\mu}_0$.  It will also have effects from  \emph{additional} correlators associated to the bulk fields $\lambda,(\Sigma+iA_5),\chi$ and also from $A_0$ directly:
\be
\Pi^{A_0}_{1}(q^2) + \Pi^{\lambda}_{1}(q^2)  =-\frac{b}{8\pi^2}\text{Ln} \left(\frac{q^2}{\mu^2}\right)+O(1/g^2_{SM})
\ee
$b= 3N$ from the adjoint source fields in the vector multiplet $V$ and $O(1/g^2_{SM})$ denotes the higher order terms  from the set of boundary operators $\{\mathcal{O}\}$ \cite{Goldberger:2002cz,Goldberger:2002hb}.  The computation of  the additional correlator $\Pi^{\lambda}_{1}(q^2)$  is very instructive.  It may be computed in either the warped or holographic picture.  In the five dimensional, or warped picture, this correlators may be computed from a bulk field.  At leading order, and below the cutoff $\mu\sim 1/L_0$, it is as if only the source field $\lambda_0$ contributes to this correlator  \cite{Goldberger:2002cz,Goldberger:2002hb}. This again confirms that one can work in either the holographic or five dimensional basis. 
It is expected that the states localised at $L_1$ will not contribute to the running of the couplings much above $E\sim 1/L_1$ \cite{Goldberger:2002cz} and for this reason one can safely ignore effects from the supersymmetry breaking sector on the IR brane.  If this is the case then perhaps this has important consequences, for instance locating an ISS type \cite{Intriligator:2006dd,Koschade:2009qu,McGarrie:2010kh} model on the IR brane one could achieve metastable supersymmetry breaking and alleviate the issue of  an early Landau pole as in this setup the messengers would not contribute to the running. 

 There are other correlators too
\be
\Pi^{(\Sigma+iA_5)}_{1}(q^2) \sim 0 , \ \ \ 
\Pi^{\chi}_{1}(q^2)\sim 0
\ee
not to mention those that arise form purely standard model fields, located on the boundary.
The gauge coupling for  $A^0_{\mu}$ may be extracted from  this collection to give
\be
\frac{1}{g^2(q^2)}=\frac{1}{g^2(\mu^2)} + \frac{b^{SSM}}{8\pi^2}\text{Ln}\left(\frac{q^2}{\mu^2}\right) -\frac{T[R] R}{g^2_5} \ \text{Ln}(qR)+...
\ee
for $\mu^2 < 1/L_0^2$.  The ellipses denote higher order terms in $g_{SM}$. The beta function coefficients of the supersymmetric standard model are given by $b^{SSM}=(-33/5,-1,3)$. These results have been used to explore unification in a warped setup \cite{Nomura:2004zs}.  

In the supersymmetric limit there are a large number of correlators which all vanish
in sets labelled by $S$:
\be
\left[3\Pi^S_1(q^2)-4\Pi^S_{1/2}(q^2)+ \Pi^S_0(q^2)\right]\equiv 0 ,\label{supertrace3}
\ee
certain collections of which give the leading order running of the four dimensional gauge coupling.  The experimentally observed correlator is in principle all orders in the hidden sector gauge couplings and, on the AdS side, can therefore be represented by arbitrary many diagrams or a sum of many correlators.    To obtain the supersymmetry breaking effects one should identify the correct contribution to the correlator and hence the relevant diagrams.

In summary, this section  has shown how supersymmetry and holography determine the UV and IR boundary conditions of the fields  and we have seen how these correspond to the same choices as those made in the Kaluza-Klein formulation of warped gauge mediation   models \cite{Goldberger:2002pc,Chacko:2003tf,Nomura:2004zs,Abel:2010uw,Abel:2010vb,McGarrie:2010yk,McGarrie:2011av,McGarrie:2012ks}.  With the condition that the supersymmetry breaking is spontaneous, it is unlikely that changing the metric by, for instance, introducing a soft wall dilaton profile \cite{McGarrie:2012ks}, will have an effect on this outcome.   As a result one may introduce new operators or new fields either in the bulk or on the IR boundary to achieve a spontaneous breaking of supersymmetry.  These new fields will play the role of messenger fields coupled to a Goldstino multiplet.

\section{Supersymmetry breaking}\label{BROKEN}
In this section we will bring together the notation and results of the previous sections to explore encoding a spontaneous supersymmetry breaking sector into the theory.   In the first instance a set of currents that encode these effects of supersymmetry breaking will be introduced, following \cite{Meade:2008wd}.  Integrating this out will generate an effective action and a set of current correlators that will then encode the effects of supersymmetry breaking. These current correlators may be used to parameterise our ignorance about the cause of the breaking.  Additionally we may imagine that there is some weakly coupled description in which the currents may be extracted from messenger fields coupled to a Goldstino multiplet.

This can be done in two ways:  the first way is to generate an effective action on the IR brane and then mediate those effects to the UV boundary. The second is to construct a four dimensional UV boundary effective action that encodes the supersymmetry breaking, at order $O(N_c^0)$ (compare with the action \refe{effectiveaction2} at order $O(N_c)$) and which couples only to external sources.  This second way corresponds to having integrated out the CFT effects.

It is useful to have a set of  criteria: as the gravity background is switched off for the moment the supersymmetry breaking theory contains an identifiable Goldstino mode and the messenger fields satisfy $\text{Str}\mathcal{M}^2=0$.  We assume that the bulk plus boundary Lagrangian is supersymmetric under the relevant supersymmetry transformations, such that the supersymmetry breaking should be a vacuum effect and therefore spontaneous.  In particular, just as the breaking of conformal symmetry is an IR effect  (in this paper implemented by the hard wall IR brane which allows for an S matrix and a discrete set of states) we also expect that the spontaneous breaking of supersymmetry is associated with the IR of the theory.   In other words the messenger fields are entirely composite degrees of freedom that should not appear in the UV.   Additionally the supersymmetry breaking currents do not directly couple to external fields: the IR brane fields couple to the CFT bulk degrees of freedom \cite{PerezVictoria:2001pa}, giving rise to form factors of vector mesons.  

 In this case we suggest locating a set of supersymmetry breaking currents coupled to a bulk gauge field  \cite{McGarrie:2010yk} on the IR brane
\be
g_{IR} \int d^5x \sqrt{-g}  \int d^4 \vartheta \mathcal{J}_{\cancel{SUSY}} V(x,z)  \delta(z-L_1) \label{brokencurrents}
\ee
using $\sqrt{-g}$ not $\sqrt{-g}_{\text{ind}}$.   In components this gives
\be
g_{IR}  \int d^5x  a^5(z) ( JD +a^{-2} j_{\mu}A^{\mu}+\lambda j +\bar{\lambda}\bar{j}  )   \delta(z-L_1).\label{unwarped}
\ee
As was discussed in \cite{McGarrie:2010yk}, to accomodate canonically normalised fields such as messenger fields from which the currents are to be extracted, one can rescale the currents 
\be
\hat{J}= a^3(z) J,  \ \   \hat{j}_{\alpha}=a^{7/2}(z) j_{\alpha},  \  \  \hat{j}_{\mu}=a^3(z) j_{\mu},
\ee
to give
\be
g_{IR} \int d^5x  ( a^2 \hat{J}D + \hat{j}_{\mu}A^{\mu}+a^{3/2}\lambda \hat{j} +a^{3/2}\bar{\lambda}\hat{\bar{j} } )   \delta(z-L_1).
\ee
These are the correct rescalings for the given metric \cite{McGarrie:2010yk}. For the moment there need not be any operator field correspondence for these currents. However it is interesting to explore constructing an operator-field correspondence for a set of  bulk messenger fields $\phi(x,z),\tilde{\phi}(x,z)$ and also for a Goldstino multiplet $X(x,z)$. Some ideas in this direction have been explored in \cite{Abel:2010vb}, for instance in constructing a bulk meson $\Phi$ which is mapped to some operator $O=Q\tilde{Q}$.  In  \cite{Abel:2010vb} the actual messengers are also located on an IR brane (they appear as magnetic quarks located on the IR brane) and so we shall find that both the soft mass formulas and the scattering cross sections to messenger fields found in the next sections will be applicable also to that model. 

\subsection{Equations of motions with currents}
Taking the equations of motion one now finds
\be
\partial_M(\sqrt{-g}g^{MN}g^{PQ}F_{NQ}(x,z))=  \hat{j}^{M}\delta(z-L_1)\label{withsource}
\ee
in which $j^{\mu}$ exists but $j^5\equiv 0$.

The dynamical part of the  D-term $D=d_{z}\Sigma$ has an equation
\be
\frac{1}{\sqrt{-g}}\partial_M\left(\sqrt{-g}g^{MN}\partial_N D(x,z)\right)-m_\Sigma^2 D(x,z)= a^{2}(z) \hat{J}\delta(z-L_1)
\ee
The symplectic Majorana spinor decomposes into a positive and negative parity spinor for which the current is associated to the positive parity component
\be
\left( g^{MN} \gamma_M D_N+m_{\Psi } \right) \Psi (x,z)=  a^{3/2}(z) \hat{j}_A\delta(z-L_1)
\ee
in which $A$ is a spinor index
\begin{equation}
j_A=\left(
  \begin{array}{c}
   j_{\alpha}\\
0
  \end{array}
\right).
\end{equation}
and the right handed current $\bar{j}_{R}^{\dot{\alpha}}$ also vanishes exactly.   One may attempt to solve these equations exactly, or apply perturbation theory by computing an effective action which give the linear response functions for such currents.

\subsection{The supersymmetry broken effective action}
In this section  the soft masses arising from spontaneous supersymmetry breaking are computed, at leading order,  for sfermions on the UV boundary and for the gaugino source that corresponds to a massless zero mode of the Kaluza Klein expansion.

\begin{figure}[ht]
\begin{center}
\includegraphics[scale=0.5]{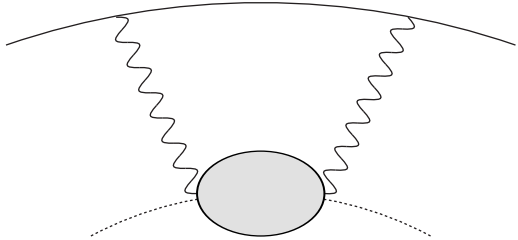}
\caption{A Witten diagram for a current correlator  of \refe{effectiveaction3}.  Operators $\mathcal{O}_{\mu}$ on the boundary are dual to gauge fields $A_{\mu}(p,z)$ which couple to a bulk current $J_{\mu}(p,z)$, which in this case is IR localised.  This diagram represents a term in the effective action on the UV boundary: $A^0_{\mu} (p)\Lambda(p) \tilde{C}_1(p^2/\hat{M}^2)\Lambda(-p)A^{0}_{\nu}(-p)$.  This contribution is $O(N_c^0)$.
} 
\label{wittendiag}
\end{center}
\end{figure}

One may specify an action located on the IR brane located at $\delta(z-L_1)$.  It will have the general form\footnote{$\sqrt{-g}$ instead of $\sqrt{-g_{\text{ind}}}$ has been used.}
\be
Z_{IR}=\sqrt{-g} \int d^5x \delta(z-L_1) \mathcal{L}_{IR}.
\ee

This can be left completely general or one may specify a perturbative action to be messenger chiral superfields, $\phi,\tilde{\phi}$, coupled to a spurion $X$  as in \refe{braneaction}.  Specifiying the perturbative action will allow us to correctly define some canonically normalised fields.

One may then integrate the generating functional on the IR brane $Z_{IR}$ and generate an effective action which now encodes spontaneous supersymmetry breaking,
\be
\delta \mathcal{L}_{eff}^{\cancel{SUSY}} |_{z=L_1}= \frac{1}{2} a^4(z)\tilde{C}(0)D^2 -ia^3(z)\tilde{C}_{1/2}(0)\lambda\sigma_{\mu}\partial^{\mu}\bar{\lambda}-\frac{1}{4}\tilde{C}_{1}(0) F_{\mu\nu}F^{\mu\nu}
\ee
\be
+\frac{1}{2}a^3(z)M\tilde{B}_{1/2}(0)\lambda\lambda+c.c.  +O(g^2_{IR}) 
\ee
at $z=L_1$.  The canonical fields are found after rescaling by \refe{rescale}.  We may define the dimensionless supertraced combination of these current correlators to be 
\be
\Omega(p^2/M^2)=\left[3\tilde{C}_1(p^2/\hat{M}^2)-4\tilde{C}_{1/2}(p^2/\hat{M}^2)+ \tilde{C}_0(p^2/\hat{M}^2)\right].
\ee
In the next subsection we apply a different approach and generate an effective action on the UV brane.

\subsection{The four dimensional action}
Due to the holographic correspondence it is always possible to relate bulk five dimensional diagrams to their corresponding four dimensional effective vertices \cite{Falkowski:2008yr}.  Typically this involves some overlap integral of bulk wavefunctions and bulk to brane propagators.  Here the situation is simplified by the exact locality of the source on the IR brane: no $\int dz$ is necessary. A similar example may be found for IR localised Yukawa couplings \cite{Gherghetta:2000qt,Contino:2004vy}.  The action for the gauge field is now written as 

\be
S_{eff}\supset \int \frac{d^4p}{(2\pi)^4}\left[\sum_SA^0_{\mu}\Pi^S_1(p^2)P^{\mu\nu} A^0_{\nu}+ g_{SM}A^0_{\mu}(\mathcal{O}^{\mu}+J^{\mu}_{SM}+\mathcal{J}^{\mu}_{\cancel{SUSY}})\right]
\ee
where the supersymmetry breaking current has been pulled from the IR to the UV
\be
\mathcal{J}^{\mu}_{\cancel{SUSY}}(p)=g_{IR}p^2\Pi_1(p^2)\Delta(p;L_0,L_1)J^{\mu}_{\cancel{SUSY}}(p,L_1)
\ee
\be
 \ \ \ \ =
g_{IR} K(p,L_1) J^{\mu}_{\cancel{SUSY}}(p,L_1) \nonumber
\ee
using the amputated boundary to bulk propagator.  This gives an effective vertex function
\be
\Lambda(p)=g_{IR} K(p,L_1)
\ee  
Taking the square of this term and Wick contracting one finds an effective term on the UV boundary

\be
S_{eff}\supset \int \frac{d^4p}{(2\pi)^4}\left[
\frac{1}{2}g^2_{SM}A^0_{\mu}P^{\mu\nu}\tilde{E}_1(p^2/\hat{M}^2)A^0_{\nu}\right]
\ee
where evidently
\be
\tilde{E}_a(p^2/\hat{M}^2)=\Lambda(p) \tilde{C}_a(p^2/\hat{M}^2)\Lambda(-p)\   \   \text{and} \ \   
M\tilde{F}_{1/2}(p^2/\hat{M}^2)=\Lambda(p) \hat{M}\tilde{B}_{1/2}(p^2/\hat{M}^2)\Lambda(-p) .
\ee
Now the set of correlators $\tilde{E}_a(p^2/\hat{M}^2)$, with $a=0,1/2,1$ are the associated ``blobs" that encode supersymmetry breaking and are order $O(N^0_c)$.

 The supersymmetry breaking effective action on the UV boundary may be written in this way:
\be
\delta L^{\cancel{SUSY}}_{eff}|_{UV}= \frac{g^2_{SM}}{2} \tilde{E}_0(0)D^2_0 -ig^2_{SM} \tilde{E}_{1/2}(0)\lambda_0\sigma_{\mu}\partial^{\mu}\bar{\lambda}_0-\frac{g^2_{SM}}{4} \tilde{E}_{1}(0) F_{\mu\nu,0}F^{\mu\nu}_0 \label{effectiveaction3}
\ee
\be
+g^2_{SM}M\tilde{F}_{1/2}(0)\lambda_0\lambda_0 +c.c.   \label{OO2}
\ee
The sources have been rescaled using \refe{rescale}.  This effective action can be represented by Witten diagrams of the type found in figure \ref{wittendiag}.   Using both the broken and unbroken UV boundary effective action all the soft mass diagrams may now be computed without reference to the fifth dimension.  This reformulates warped gauge mediation models in a way much closer to the original GGM proposal \cite{Meade:2008wd}.

One may now compute various soft terms using the four dimensional boundary action.  The formula for sfermion masses is now given by
\be
m^2_{\tilde{f}}=-g^4_{SM}\int \frac{d^4p }{(2\pi)^4}\frac{1}{p^2}\left[\Lambda^2(p) \Omega(p^2/\hat{M}^2)\right]
\ee 
Each diagram that contributes to the result above is made of two bulk to boundary propagators and then a current correlator located at $z=L_1$.  The function $\Omega$ contributes a negative sign and the sfermion masses are positive \cite{Meade:2008wd}.
The factors of $g^4_{SM}$ arise from \refe{1}, \refe{currentsource} and \refe{brokencurrents} and appears to carry an extra factor of $g^2_{IR}$.   Motivated by current field identities  \cite{McGarrie:2012ks} we may wish to think of  $(g^2_{SM}/g_{\rho}) V^0\mathcal{O}$ in \refe{currentsource}  and then $g^2_{SM}(g^2_{SM}/g_{\rho})g^2_{IR}$ would appear, to give the right overall $g^4_{SM}$.  We expect there is a natural resolution to this and keep the factors of $g$ explicitly. 

The next contribution to sfermion masses is 
\be
\delta m^2_{\tilde{f}}=g^4_{SM}g^2_{IR}\int \frac{d^4p }{(2\pi)^4}\frac{1}{p^2}\Lambda^2(p) \Delta(p,z,z) \left[\hat{M}\tilde{B}_{1/2}(p^2/\hat{M}^2)\right]^2
\ee
evaluated at $z=L_1$ and is pictured in figure \ref{wittendiag3}.  Just as the discussion above this should naturally appear with a factor $g^6_{SM}$ with a factor $g^2_{IR}$ inside the effective vertices. The Green's function carries a factor $\frac{12\pi^2}{N_c}$.  There is clearly some tension between the leading and subleading diagrams as the second is suppressed by $1/N_c$ or a loop factor, however for a messenger model the first diagram will be found to be suppressed by $y^2\sim|\hat{M}L_1|^2$. For an $N_c\sim 10$  would have to be  $y\sim 0.3$ for both diagrams to be of the same order.  This diagram is most relevant in the ``Gaugino Mediated'' limit \cite{Kaplan:1999ac,Chacko:1999mi,Schmaltz:2000ei}.

\begin{figure}[ht]
\begin{center}
\includegraphics[scale=0.5]{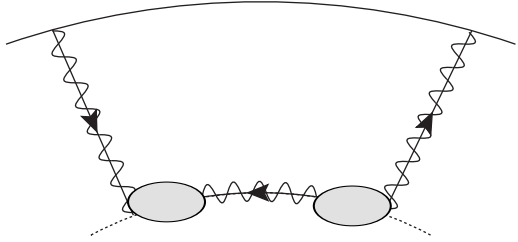}
\caption{A Witten diagram for a subleading contribution to the scalar soft masses from a double mass insertion of the gaugino soft mass. The operator $\mathcal{O}_{\alpha}$ on the boundary is dual to the field $\lambda_{\alpha}$ for which $\lambda^0$ is the source.   This soft mass is $O(1/N_c)$.
} 
\label{wittendiag3}
\end{center}
\end{figure}
It is interesting to explore this contribution from the perspective of the four dimensional boundary action.   Using the identity
\be
\Delta(p;z,z')=\Delta(p;L_0,L_0)K(p,z)K(p,z') + D(p;z,z')
\ee
where $D(p,z,z')$ is a Dirichlet Green's function, one finds 
\be
\delta m^2_{\tilde{f}}=g^4_{SM}\int \frac{d^4p }{(2\pi)^4}\frac{1}{p^4\Pi(p^2)} \left[M\tilde{F}_{1/2}(p^2/M^2)\right]^2 + ...
\ee
where the ellipses signify a piece proportional to the Dirichlet Green's function which vanishes on the IR brane. This is  what one would have found using only the four dimensional effective action \refe{effectiveaction3} as effective vertices.  Specifically one takes two insertions of the Majorana soft term in \refe{effectiveaction3}.

The gaugino mass to the sources  $\lambda^0$ appear to take the form
\be
m_{\lambda}=\lim_{p^2\rightarrow 0 }g^2_{SM} \Lambda^2(p) M\tilde{B}_{1/2}(p^2/M^2) \label{not}
\ee
and we expect the cross sections associated with these current correlators will take this form.  However
\be
 \lim_{p^2\rightarrow 0}\Lambda(p^2)=g^2_{IR}
\ee
as a typical form factor $F(0)=1$, so this soft term appears with an extra factor of $g^2$.  If we compare this with the Kaluza Klein expansion  of $\lambda(x,z)$  one would expect a term directly as 
\be
m_{\lambda}\lambda_n \lambda_m=g^2_{IR} M\tilde{B}_{1/2}(0)\lambda_n \lambda_m
\ee
on the IR brane. 

Higher order, subleading in $g_5$, corrections may be computed  following \cite{Lee:2010kb} for both sfermion and gaugino masses.  The final mass eigenstates are a mixture of the soft breaking Majorana masses and the Dirac Kaluza-Klein masses which may be found following \cite{Marti:2001iw,ArkaniHamed:2001mi}.    A Casimir energy contribution to the vacuum energy will arise due to the breaking of supersymmetry \cite{McGarrie:2010kh,McGarrie:2010yk}.  Including also supergravity contributions it may be used to stabilise the brane separation and fix the value of $R$.

To summarise, the procedure for computing soft masses by using a four dimensional boundary action with the bulk to boundary propagator $K(p,z)$ encoded into an effective vertex $\Lambda(p)$ and then dividing (canonically normalising the sources $V^0$) by $\Pi_a$ is equivalent to  computing the soft masses using the bulk Green's functions as in warped models \cite{McGarrie:2010yk}. This approach gives considerable insight into the  four dimensional interpretation of warped models of general gauge mediation.
\subsection{Soft masses for a messenger model}
In the previous section we have accomplished the primary objective of integrating out the bulk theory to create a UV boundary action that encodes supersymmetry breaking, with dressed vertices associated with the intermediate CFT states.  However we have in mind that a supersymmetry breaking sector and messenger fields arise on the IR brane \cite{Abel:2010vb}, closely related to the magnetic description of ISS models \cite{Intriligator:2006dd}.  It is therefore useful to give formulas for their soft masses and also to compute cross sections to these states.

To do this we introduce a simple O'Raifeartaigh model on the IR brane which is straightforward to generalise \cite{Marques:2009yu}:
\be
\mathcal{L}_{IR}=   \left(\int d^4\theta  a^3(L_1)\phi^{\dagger}e^{2V}\phi +a^3(L_1)\tilde{\phi}^{\dagger}e^{-2V}\tilde{\phi}  \right)  +\left(\int d^2\theta a^4(L_1)W + c.c.\right)\label{braneaction}.
\ee 
It is from the kinetic terms of these fields that the supersymmetry breaking curents are extracted.  This action suggests that one then identify $g_{IR}$ associated with the gauge groups in the action \refe{braneaction}.  Further one may interpret these messenger fields as composite fields that live in the IR. Taking a simple superpotential
\be
W=X\phi\tilde{\phi}
\ee
with $X=M+\vartheta^2 F$. $M$ is the characteristic mass scale of the hidden sector, or messenger mass scale and $F$ is the F-term of spontaneous symmetry breaking. To determine the correct rescalings associated with canonical normalisation of the kinetic terms,  we choose to rescale 
\be
\hat{M}=a(z)M, \ \  \hat{F}=a^2(z) F.
\ee
The components of the canonically normalised messenger fields $\hat{\phi}_{\pm}$ have masses
\be
m^2_{\pm}=\hat{M}^2 \pm \hat{F} .
\ee
Using \refe{braneaction} and the appropriate limits found in \cite{McGarrie:2010kh} we determine
the gaugino soft mass to be 
\be
m_{\lambda}=\left(   \frac{\alpha_{IR,r}  }  {4\pi}    \right)\left(\frac{R}{z}\right)_{z=L_1}\sum_{i=1}^p\left[\frac{d_r(i)F}{M}2g(x_i)\right],
\ee
where $p$ is the number of messengers and $d_r(i)$ is the Dynkin index for the i messenger in the group $r$, where $r$ labels the standard model gauge groups of the weakly gauged flabour symmetry $SU(N)_F\supset U(1)\times SU(2)\times SU(3)$.  The function $g(x_i)$ can be found in \cite{Martin:1996zb} and for $x<1$ $g(x)\sim1$.  In the limit $m_{\rho}\ll M$
 the external sfermion masses are found to be 
\be
m^2_{\tilde{f}}\sim \sum_r C^r_{\tilde{f}} \left(  \frac{\alpha_{SM,r}  }  {4\pi}    \right)^2 
\left(\frac{R}{z}\right)^2_{z=L_1}\sum_{i=1}^p 2d_r(i) |\frac{F}{M}|^2|\frac{1}{\hat{M}}|^2 
 \int d p  \ p  \ \Lambda^2(p)
\ee
where the relevant quadratic Casimir is labelled $C^r_{\tilde{f}}$.   The integral on $\Lambda(p)$ will supply a scale $m_{kk}$ after change of variables and likely also a factor $(R/L_1)$. Whilst the limit $m_{kk}\sim \hat{M}$ is not possible to compute analytically for such a complicated model we know from simpler models that there is also a hybrid regime \cite{Auzzi:2010mb,McGarrie:2011dc} as well.  The hybrid regime is more promising for phenomenology however numerical methods would be required to determine this regime from these warped/holographic models.  It is useful to discuss how the scale $M$ is determined. In Randall-Sundrum \cite{Randall:1999ee} models IR scales such as $M$ are naturally of the order $M_{Planck}$ and then $\hat{M}=Me^{-kl}$ sets a TeV scale.  However $\hat{M}$ in our case is essentially the vev of a scalar $\phi_G$  component of Goldstino multiplet.  It is a classical modulus and get is vev from the Coleman Weinberg potential with contributions from the same scale that gives mass to the magnetic gauge bosons: $m_{kk}\sim 1/L_1$ \cite{Abel:2010vb}. One therefore expects $|\hat{M}L_1|\sim O(1)$ is possible and overlaps with four dimensional constructions \cite{Green:2010ww}.

\section{Scattering}\label{SCATTERING}
\begin{figure}[ht]
\begin{center}
\includegraphics[scale=0.5]{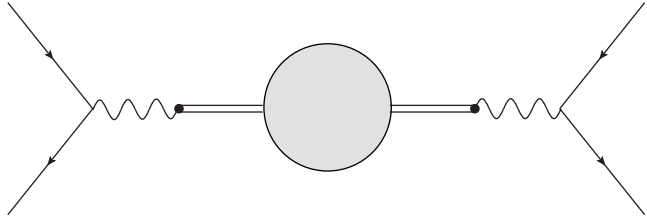}
\caption{A diagram to represent certain pieces contained in the full matrix element $i\mathcal{M}(e^+e^- \rightarrow e^+e^- )$.
One 
applies the optical theorem to obtain $\sigma( e^+e^- \rightarrow hidden)$.
The double lines denote
intermediate meson resonances. The blob denotes
the current correlator $\tilde{C}_1(s)$ in \refe{scattering1}.
The cut is applied across the correlator.   This diagram is equivalent to the Witten diagram in  figure \ref{wittendiag}.
} 
\label{eplusdiagram}
\end{center}
\end{figure}
We now wish to discuss scattering cross sections $\sigma_a(visible \rightarrow hidden)$ with intermediate resonances \cite{Fortin:2011ad,McGarrie:2012ks}.  If one integrates out the full AdS system one could obtain the full current correlators analogous to \refe{full5}.  The correlator $\Pi_{full}(q^2)$ may be thought of as the correlator, ``to all orders'', after fully integrating out the bulk and IR brane.  $\Pi_{full}(q^2)$ contains many pieces associated to the sum over all intermediate states.   For instance taking a cut across figure \ref{wittendiag0}, which is contained in \refe{full5}, would give the cross section $\sigma(vis\rightarrow A_n)$  where $A_n$ are the vector meson kk states of the bulk field $A_{\mu}(x,z)$.

 However, we are primarily interested in the matrix elements contained in $\Pi_{full}(q^2)$ that will give $\sigma(visible\rightarrow hidden)$  where \emph{hidden} actually stands for the supersymmetry breaking subset of the hidden sector: the states  contained in the currents $\mathcal{J}_{\cancel{SUSY}}$.  This subset of matrix elements are pictured in figure \ref{eplusdiagram}.  The ``blob'' in figure \ref{eplusdiagram} represents the current-current correlator of $J^{\mu}_{\cancel{SUSY}}$.

  Just as in AdS/QCD models one obtains a form factor of vector mesons, which in this setup is essentially associated with the boundary to bulk propagator of the approximate CFT or equivalently the amputated Green's function $\Delta(p;z,z')$ of the AdS description.

The expression for the scattering cross sections is given by 
\be
\sigma_a(s,vis\rightarrow hid)=\frac{ (4\pi\alpha_{SM})^2}{2s} \Lambda^2(s)\ \text{Disc}  \ \tilde{C}_a(s/\hat{M}) \label{scattering1}
\ee
We wish to rewrite this result in terms of AdS/QCD.  Using the identifications of  appendix \ref{scatteringgreen} one may finally 
 arrive at a general expression for the scattering cross sections
\be
\sigma_a(vis\rightarrow hid)=\frac{(4\pi\alpha_{SM})^2}{2s} \left(g^2_{IR}g^2_5\right)\sum_{n=1}\frac{F_n\psi_n(z)  }{p^2+m_n^2} \sum_{\hat{n}=1}\frac{F_{\hat{n}} \psi_{\hat{n}}(z) }{p^2+m_{\hat{n}}^2} \ \text{Disc}  \ \tilde{C}_a(p^2/\hat{M}) \label{scattering1}
\ee
taking $z=L_1$ and where the correlators $\tilde{C}_a(s)$ have an explicit form, if one specifies a messenger sector, which may be found in \cite{Fortin:2011ad}.   The scattering cross section is at $O(N^0_c)$  due to the factors  $g^2_5  \ F_n \  F_{\hat{n}}$.  

These form factors have a natural interpretation \cite{Erlich:2006hq} in terms of Migdal's Pade approximation of the OPE.  In particular the form factors are very similar to those found in AdS/QCD models.   It encodes a sum of monopole contributions of an infinite tower of vector mesons with decay constants $F_n$ for each meson. 
 It is useful to understand the  difference however.  Normally for a bulk current that is identifiable there is an overlap integral between the non normalisable and normalisable modes in the effective vertex
\be
\Lambda_{s}(Q)=\int dz a^s(z) \mathcal{J}(Q,z) \varphi(z)\tilde{\varphi}(z)
\ee
where for instance $\varphi(z),\tilde{\varphi}(z)$ are the wavefunctions of the fields in the current.  This could then be rewritten into the definition of the coupling constant $g_n$ analogous to $g_{\rho\pi\pi}$:
\be
g_n=g_5 g_{IR}\int dz \psi_n(z)  \varphi(z)\tilde{\varphi}(z) \delta(z-L_1).
\ee
 As the currents in this example are IR localised there is no bulk wavefunction and further $\int dz \delta(z-L_1)$ results in no integral overlap.  These are considerable simplifications and it may be desirable to have bulk messenger fields to achieve these additional features. This would  require a more exotic messenger sector such as bulk hypermultiplets.

Let us discuss vector meson dominance in such a model  \cite{Hong:2004sa,Hong:2005np,RodriguezGomez:2008zp}.  The messenger fields are contained in the same current multiplet and similarly the vector mesons and mesinos in the same vector multiplet such that the locality of currents in the IR and supersymmetry of the bulk action fixes for each $n$ 
\be
g_{A \phi \tilde{\phi}}=g_{A\psi \tilde{\psi} }= g_{\lambda \phi \tilde{\psi}}= g_{V\Phi\tilde{\Phi}}.
\ee
Interestingly this result is related to the effective vertex $\Lambda(p)$ and the supersymmetry breaking effective action on the UV brane \refe{effectiveaction3} which is UV convergent.  

 These $g_n$ are not related to $g_{AAA}$ , $g_{A\lambda\bar{\lambda}}$ and $g_{A\Sigma\Sigma}$ which are themselves related through supersymmetry of the Yang Mills bulk action and do involve an integral of the form   $\int dz \psi_{\hat{\hat{n}}}(z)\psi_{\hat{n}}(z)\psi_n(z)$.  Again a closer similarity between couplings of vector-vector-vector to vector-matter-matter may be achieved by locating the messenger fields in the bulk and taking account of their bulk profiles. So it seems that one is unable to get a fully universal set of relations between couplings that one may find in similar constructions that start with the magnetic description of SQCD \cite{Komargodski:2010mc,Abel:2012un} and may be  a prediction to descriminate between the two cases.  It would be nice if there was a natural way to integrate out some of the bulk degrees of freedom and arrive at a quiver-like model similar to those explored in \cite{Auzzi:2010mb,Auzzi:2010xc,Auzzi:2011wt} somewhat in the direction of \cite{Abel:2010vb}.

We may also define the size of the messenger field's charge distribution from the form factor through
\be
\braket{r^2}=6\frac{\partial}{\partial p^2}F(p^2)|_{p^2=0}=  -\sum_n\frac{6g_5g_{IR}^2F_n\psi_n(L_1)}{m^4_n}.
\ee
  Taking $\psi_n(L_1)=C(-1)^n$,  $F_n=F$ the decay constant (not to be confused with the F-term of supersymmetry breaking) and $m_n^4 = m^4_{\rho} n^4$ one obtains
\be
\braket{r^2}=\frac{42 g_5 g_{IR}^2F C \pi^4 }{720m^4_{\rho}}\sim O(1/m_{\rho}^2).
\ee
As $g_5 F_n =O(N_c^0)$ we see that the effective radius does not depend on $N_c$.    This may be compared with the simple two site quiver model arising from a magnetic description of SQCD  \cite {McGarrie:2012ks,Komargodski:2010mc,Auzzi:2010mb} in which
\be
\braket{r^2}=\frac{6}{m^2_{\rho}}.
\ee
It may be worthwhile to consider possible sum rules that can be derived from these results. 

We also wish to make a final observation with regard to these cross sections.  Including all contributions to the correlator, both vector meson hidden states and susy breaking hidden sector states, we would finally obtain a full $\Pi(s)_{hidden}$.  One might then hope to define a duality in $e^+,e^- \rightarrow$  hidden states analogous to Sakurai's duality in $e^+,e^- \rightarrow$ hadrons \cite{Sakurai:1973rh}.  Whilst in that case the vector meson spectrum was Regge-like, with $1/s$ scaling, still we can imagine that by taking integrals of the cross sections over a range $\Delta s$, where  $\Delta s/\sqrt{s}\gg 1/L_1$ holds, i.e. a sort of smearing procedure, we might be able to match integrals of the cross sections to some perturbative SQCD description, reproducing features of the duality.
\section{Conclusions and discussion}\label{CONCLUDE}
In this paper we have continued the programme of exploring spontaneous symmetry breaking in five dimensional models using the framework of general gauge mediation \cite{Meade:2008wd,McGarrie:2010yk,McGarrie:2011dc,McGarrie:2012ks,McGarrie:2010kh,McGarrie:2010qr,McGarrie:2011av}.  In particular this is a holographic model based on AdS/QCD, where previously spontaneous supersymmetry breaking models have focused on the five dimensional aspect using a Kaluza-Klein basis  \cite{Goldberger:2002pc,Chacko:2003tf,Nomura:2004zs,Abel:2010uw,Abel:2010vb,McGarrie:2010yk,McGarrie:2011av,McGarrie:2012ks}.  Whilst of course the Kaluza-Klein and holographic picture overlap significantly, it is insightful to explore both perspectives. This construction allows for an entirely UV localised four dimensional effective action that encodes not only supersymmetry breaking but also encodes the full effective vertex corrections due to the intermediate CFT states. We would like to interpret the messenger sector and spurion  of this model as arising from a dynamically broken metastable supersymmetry breaking sector \cite{Intriligator:2006dd} as part of an approximate CFT.  This description has given a more natural encoding of scattering cross sections \cite{Fortin:2011ad} to these states.

Some of the key results of this paper are that we have been able to identify a natural interpretation of the terms in the effective action as an expansion in $N_c\sim 1/g^2_5$.  Using an entirely four dimensional effective action on the UV boundary, the effects of spontaneous supersymmetry breaking has been encoded and various soft masses computed. In addition one is  able to identify the form factor for scattering cross sections in close analogy to AdS/QCD models.

To develop this further, it would be interesting to introduce some \emph{bulk} messengers and Goldstino fields in analogy to the pion field $X(x,z)$ of AdS/QCD, which is dual to $\bar{q}q$ on the boundary.   This would allow us to identify an operator dual to these bulk messenger fields, from which the bulk supersymmetry breaking current may be extracted. Perhaps a model  of this nature could be achieved by first starting with maximal super Yang Mills in five dimensions and taking an orbifold, thereby generating additional bulk adjoint multiplets \cite{McGarrie:temp}.  Also motivated by AdS/QCD models, 4 point functions have been computed \cite{Hambye:2006av} and gravity to hadron form factors which involve metric perturbations.  Indeed it would be interesting to compute the supercurrent correlator on the IR brane $\braket{S^{\mu}_{\alpha}S^{\nu}_{\beta}}$ following \cite{McGarrie:2012dg}. These would be interesting both in terms of extending the scattering programme of this paper as one expects form factors to appear, but also to understand its embedding in supergravity more concretely. The framework of this paper will have a natural extension in terms of  the soft wall model \cite{Karch:2006pv}, where a more Regge like trajectory in which mesons scale as $m^2_n\sim  n m^2_0 $ instead of $n^2$, which will effect form factor in scattering cross sections.   One could also include both $SU(N_f)_L\times SU(N_f)_R $ flavour groups in the bulk perhaps to explore the addition of chiral symmetry breaking, where this paper has in some sense looked simply at  the diagonal $SU(N_f)_D$.  Additionally one might expect that the CFT dual to this model has baryonic states: they would need to be mapped to solitons of this AdS setup.

    An interesting observation \cite{Bhattacharyya:2012qj}  that the addition of  an adjoint chiral superfield of the $SU(3)$ sector may assist in a 125GeV Higgs within an extended standard model framework.  In addition Dirac soft masses \cite{Benakli:2008pg}  may arise naturally from the interaction of the fermionic component of an adjoint chiral superfield and the vector superfield fermion.  Within the framework of this paper one would require a $\Phi_{adj}$ to obtain positive parity on both branes.  There are two immediate concerns: the value of the vev of scalar component so as not to Higgs the standard model gauge groups and the procedure for determining the boundary action with coupled fermions as in section \ref{BOUNDARY}.  Perhaps this may be obtained starting from maximal super Yang Mills in five dimensions: we hope to say more in \cite{McGarrie:temp}. It is also particularly interesting that the method of computing a current current correlator overlap significantly with the methods used in AdS/condensed matter applications \cite{Faulkner:2010zz}.  With this in mind it may be interesting to explore this type of construction at finite temperature.

\paragraph{Acknowledgements} 
MM is funded by the Alexander von Humboldt Foundation.  MM would like to thank Andreas Weiler, Daniel C. Thompson, Felix Brummer and Alberto Mariotti for fruitful discussions.  The diagrams are drawn in JaxoDraw  \cite{Binosi:2003yf,Binosi:2008ig}.


\appendix

\section{The response to a bulk source}\label{BULKFIELDS} 
In this appendix  wish to understand the response of a bulk field to these source currents.  This will determine for us both the bulk to boundary propagator and the Green's function.  One can then apply the particular case where the bulk field is localised on the IR brane, $J^{5d}=J^{4d}\delta(z-L_1)$, without any difficulty.  These results are necessary for the computations in section \ref{BROKEN}.

In the literature there are a multitude of  forms that the Green's functions can take \cite{Randall:2001gb,Randall:2001gc,Goldberger:2002hb,Gherghetta:2000kr,Pomarol:2000hp,PerezVictoria:2001pa,Csaki:2010aj} and also limits explored in \cite{Randall:2001gb,Randall:2001gc,Ichinose:2006en,Ichinose:2007zz}.
This is in part due to euclideanisation $q=ip$ and also as there are various relations between the Bessel functions $J,Y$ and modified Bessel functions $H,K,I$.  There are also a number of possible choices of gauge.

\subsection{Holographic renormalisation}
The results of this paper are quite general and may be applied to any holographic construction, by following the prescription of holographic renormalisation following \cite{Skenderis:2002wp,PerezVictoria:2001pa}. 

We wish to introduce some boundary source fields analogous to the canonical example $\phi_0$: $(A^{\mu}_0,\lambda^{\alpha}_0,D_0 )$. Next we introduce a set of bulk operators, (analogous to $\frac{1}{3}\Phi^4$ \cite{Skenderis:2002wp} and in particular page 23, for instance.)  which may be extracted from fields coupled to the supersymmetry breaking and label these $(J_{\mu},J_{\alpha},J)$.  Following  \cite{Skenderis:2002wp} we wish to solve for the bulk fields perturbatively in $g_5$. Writing  out explicitly
\be
 \ \ A^{\mu}=A_{free}^{\mu}+g_5 A^{\mu}_1+\cdots \label{Aexpanded}
\ee
\be
 \ \ \lambda^{\alpha}=\lambda_{free}^{\alpha}+g_5 \lambda^{\alpha}_1+\cdots
\ee
\be
 \ \  D=D_{free}+g_5 D_1+\cdots
\ee
The ellipses denote higher order corrections.
The fields denoted $\Phi_{free}$ are solutions of the free equations of motion and those of $\Phi_1$ are of the equations of motion with sources (as in \refe{withsource}).  Quite generally, the free solutions are solved by 
\be
A^{\mu}_{free}(x,z)= \int d^4y K(z, x-y) A^{\mu}_0(y)\label{freesource}
\ee
Where $K(x-y;z)$ is the bulk to boundary propagator in position space.   The bulk field's response to a bulk source \cite{Gherghetta:2000kr,Pomarol:2000hp,PerezVictoria:2001pa,Csaki:2010aj} is given by
\be
A^{\mu}_1(x,z)=\int d^4y dz' \sqrt{-g}\Delta(x,z:y,z')    a^{-2}(z)  j^{\mu}(y,z')    \label{sourcecurrents}
\ee
where $j_{\mu}(y,z)$ is a bulk current  and $\Delta(x,z:y,z')$ is the Green's function,  $\sqrt{-g}$ is the d+1 dimensional metric.  The factors also agree with \refe{unwarped}. A boundary current effects the on shell source field as 
\be
A_{0, \mu}(x)=\int d^4y \sqrt{-g}\Delta(x,L_0:y,L_0) a^{-2}(z) j_{\mu}(y) .
\ee
Hence one may define two objects, the Green's function and the bulk to boundary propagator.  The bulk  Green's function is found from \cite{Randall:2001gb,Gherghetta:2000kr,Csaki:2010aj}, by first making the identification
\be
\Delta(x,z; x',z')=\int \frac{d^4p}{(2\pi)^4}e^{ip.(x-x')}\Delta(p;z,z').
\ee
  One can invert the supersymmetric correlators \refe{pi1}  to define the Green's functions \cite{Gherghetta:2000kr,Abel:2010vb} such that (taking $q=ip$)
\be
\Delta(p;z,L_0)=\frac{1}{q^2}\frac{qg_5^2}{a(L_0)}  \frac{ \left[Y_0(q z) J_1(q L_0)-J_0(q z)Y_1(q L_0) \right] }{ \left[Y_0(q z) J_0(q L_0)-J_0(q z)Y_0(q L_0) \right]}
\ee
The equivalence of the holographic and Kaluza-Klein basis is made by the  identification
\be
\int dz \int dz' \Delta(p;z,z')=\frac{g^2_5}{R}\sum_n \int d r \int dr' \frac{f_n(r)f_n(r)}{p^2-m_n^2}.
\ee

 Taking a Fourier transform of \refe{freesource} one arrives at
\be
A_{\mu}(p,z)= A^{0}_{\mu}(p)  K(p,z).
\ee
This determines the bulk to boundary propagator  \refe{bb} up to a normalisation  
\be
K(p,z)= \frac{V(p,z)}{V(p,L_0)}\label{gtilde}
\ee
where $V(p,L_0)=1$ is often used \cite{Erlich:2005qh,Erlich:2006hq}.   This naturally relates the bulk Green's function and bulk to boundary propagator through
\be
\Pi_1(p)= -\frac{1}{p^2 g_5^2} a(z)\partial_z K(p,z)|_{z=L_0}
\ee
which when inverted gives
\be
\Delta(p,L_0,L_0)=\frac{1}{p^2}\Pi_1(p)^{-1}=- g^2_5\left[a(z)\partial_z K(p,z)\right]^{-1}|_{z=L_0}.
\ee
It is also useful to define
\be
K(p,z)=p^2\Pi_1(p^2) \Delta(p,L_0,z)=\frac{ \Delta(p,L_0,z)}{ \Delta(p,L_0,L_0)}
\ee
which is the amputated boundary to bulk propagator.  Conversely one may write
\be
\Delta(p,L_0,z)=g^2_5 \frac{K(p,z)}{\left[a(z)\partial_z K(p,z)\right]|_{z=L_0}}.
\ee 
In summary one may work with either bulk Green's functions or the boundary to bulk propagator.  Typically one finds that holographic models compute similar diagrams but with an amputated Green's function, compared with warped constructions.

\subsection{Pulling operators from the IR to the UV}
In the holographic picture one may first integrate out the bulk theory and generate an effective boundary Lagrangian.   Let us explore a few useful examples.  The first \cite{Contino:2004vy,Davoudiasl:2009cd} example is an IR localised Yukawa 
\be
\int d^5x \delta (z-L_1) \sqrt{-g_{\text{ind}}}\left[y_5 q \phi \Psi \right].
\ee
The physical Yukawa is then generated by pulling the coupling to the UV brane with a Green's function and then amputating that Green's function by the UV to UV Green's function:
\be
y= y_5\frac{ \Delta(p,L_0,L_1) }{\Delta(p,L_0,L_0)}\mathcal{N}=y_5 K(p,L_1) \mathcal{N}
\ee
where $\mathcal{N}$ is a normalisation $a(L_1)^{3/2}/\sqrt{Z_0(\mu)}$ for the fermions.  Let us look at another example.  For a general bulk current $J_{\mu}(p,z)$  one has a coupling 
\be
A_{\mu}(x,z)J^{\mu}(x,z)
\ee
To write this as a UV boundary operator one connects with a Green's function and then again amputates
\be
\int d z A^0_{\mu} \frac{\Delta^{\mu\nu}(p;L_0,z) }{\Delta(p,L_0,L_0)}  J_{\nu}(p,z)=\int d z  K(p,z) A^0_{\mu}(p)J^{\nu}(p,z)
\ee
In general this defines for us an effective vertex on the UV boundary 
\be
\Lambda^{\mu}(p)= \int dz K(p,z)  J^{\mu}(p,z).
\ee
In the particular case that $J(p,z)= J(p)\delta(z-L_1)$ one obtains
\be
g(p)A^0_{\mu}(p)J^{\mu}(p)=K(p,L_1)A^0_{\mu}J^{\mu}
\ee
where 
\be
p^2\Pi_1 (p^2) \Delta(p,L_0,z)=K(p,z).
\ee
So it appears that the main distinction between the warped models \cite{McGarrie:2010yk} and this holographic model is that one may additionally amputate by the UV boundary to UV boundary Green's function.  This example will also allow us to generate a UV boundary effective action that encodes the supersymmetry breaking effects.

\subsection{Green's functions for scattering}\label{scatteringgreen}

The natural relationship between the bulk to boundary propagator and the bulk Green's function will allow one to compute effective one loop diagrams in the bulk theory.  It appears in the AdS/QCD literature \cite{Erlich:2005qh,Karch:2006pv} that for computing form factors one chooses another basis for the Green's function:
\be
p^2\Pi_1(p)= -\frac{1}{p^2 g_5^2} a(z)\partial_z \tilde{G}(p,z)|_{z=L_0}=\frac{1}{ g_5^2}\left[ a(z)\partial_z[a(z)\partial_z G(p,z,z')\right]|_{z=z'=L_0}
\ee
One may solve the equations of motion of \refe{eom1} and \refe{eom2}  with this Green's function
\be
G(p;z,z')=\sum_{n=1} \frac{\psi_n(z) \psi_n(z')}{p^2-m^2_n+i\epsilon}.\label{greensfunction}
\ee
These are not the same as the KK expansion, by using $p^2=M_n^2$ and satisfying $\psi_n(0)=0$ and $\partial_z\psi_n(L_1)=0$.    This expansion does not include the massless sources. 
The decay constants may now be related to the $\psi_n(x)$ functions
\be
F_n=\frac{1}{g_5}\left[\left(\frac{R}{z}\right)\partial_z \psi_n(z)\right]_{z=L_0} \ \  \text{and} \ \   F'_n=\frac{1}{g_5}\left[\left(\frac{R}{z}\right)\partial_z \psi_n(z)\right]_{z=L_1}.
\ee
which can be determined at tree level from
\be
F^2_n=\lim_{p^2\rightarrow m_n^2}\left[(p^2-M_n^2) \Pi(p^2) \right].
\ee
Following \cite{Grigoryan:2007vg} the decay constants are defined as 
\be
F_n\epsilon_{\mu}=\braket{0|\mathcal{O}_{\mu}|\rho_n}, \ F_n=\braket{0|\mathcal{O}|\rho_n}  \ \text{and}  \  F_n\epsilon_{\alpha}=\braket{0|\mathcal{O}_{\alpha}|\rho_n}.
\ee
We have suppressed the group indices $\mathcal{O}^a T^a$, $\rho^b$ delivering a $\delta^{ab}$.
As in \cite{Grigoryan:2007vg}  one takes
\be
K(p,z)=\frac{V(q,z)}{V(q,L_0)}= - g_5\sum_{n=1}^{\infty}\frac{F_n\psi_n(z)}{p^2-m_n^2}
\ee
Taking $q^2=-Q^2$ the non normalisable mode is given by 
\be
\mathcal{J}(Q,z)=Qz\left[K_1(Qz)+I_1(Qz)\frac{K_0(QL_1)}{I_0(QL_1)}\right]= g_5\sum_{m=1}^{\infty}\frac{F_m\psi_m(z)}{Q^2+m_m^2}.
\ee
These identities  will be used to compute scattering cross sections in section \ref{SCATTERING}.

\section{Notation and Conventions}\label{APP1}
We label the five dimensional indices $M,N$ and $A,B$ with four dimensional indices $\mu,\nu$. We use a  mostly positive metric $\eta_{\mu\nu}=\text{diag}(-1,1,1,1)$.  The Dirac algebra in a curved geometry is given by 
\be
\{ \Gamma^M,\Gamma^N\}=2 G^{MN}
\ee
through the relation $\Gamma^M=e^M_A\gamma^A$ we may find some locally flat coordinates such that 
\be
\{ \gamma^M,\gamma^N\}=2 \eta^{MN}.
\ee
In this case $e^{M}_{A}= \delta^M_A/a(z)$.  The covariant derivative acting on spinors is given by 
\be
D_M=\partial_M + \frac{1}{8}\omega_{M AB} [\gamma^{A},\gamma^{B}]
\ee
The spin connection is related to the gamma matrices by
\be
\omega_{\mu a 5}=\eta_{\mu a}/a(z)\partial_z a(z).
\ee
In two component spinor notation
\be
\gamma^M=\left(\,\left(\begin{array}{cc}0&\sigma^\mu_{\alpha \dot{\alpha}}\\ 
\bar{\sigma}^{\mu \dot{\alpha} \alpha }&0
\end{array}\right),
\left(\begin{array}{cc}-i&0\\ 0&i\end{array}\right)\,
\right)\,,~~\mbox{and}~~~
C_5=\left(\begin{array}{cc}
-\epsilon_{\alpha\beta} & 0\\ 
0 & \epsilon^{\dot\alpha \dot\beta}
\end{array}\right)\,,
\ee
where $\sigma^\mu_{\alpha \dot{\alpha}}=(1,\vec{\sigma})$ and
$\bar{\sigma}^{\mu \dot{\alpha}
\alpha}=(1,-\vec{\sigma})$. $\alpha,\dot{\alpha}$ are spinor indices
of $\text{SL}(2,C)$.   The $\gamma^4_{5d}=-i \gamma^5_{4d}$  where explicitly
\be
\gamma^5_{4d}=\left(\begin{array}{cc}
-\mathbf{I}& 0\\ 
0 &\mathbf{I}
\end{array}\right)\,.
\ee

\bibliographystyle{JHEP}
\bibliography{soft.bib}

\end{document}